\documentclass[12pt]{article}
\usepackage{amsmath,amssymb,graphicx,amsfonts}
\usepackage{multicol,color,longtable}
\usepackage{mathtools}
\usepackage{latexsym}

\usepackage[nosort]{cite}

\usepackage{hyperref}

\numberwithin{equation}{section}

\newcommand{\bit}{\begin{itemize}}
\newcommand{\eit}{\end{itemize}}
\newcommand{\beq}{\begin{equation}}
\newcommand{\eeq}{\end{equation}}
\newcommand{\bea}{\begin{eqnarray}}
\newcommand{\eea}{\end{eqnarray}}

\definecolor{darkred}{rgb}{0.65,0.15,0}
\hypersetup{pdfborder={0 0 0},colorlinks=true,urlcolor=blue,citecolor=blue,linkcolor=darkred,linktocpage=true}

\newcommand{\dd}{\mathrm{d}}

\def\GN{{\rm G}_{\rm N}}

\def\p{\partial}
\def\d{\delta}
\def\L{\Lambda}

\def\nn{\nonumber}
\def\eps{\epsilon}
\def\ve{{\tilde{\varepsilon}}}

\def\cC{{\cal{C}}}

\def\cF{{\cal{F}}}
\def\cG{{\cal{G}}}
\def\cH{{\cal H}}
\def\cN{{\cal N}}

\def\cP{{\cal{P}}}
\def\cO{{\cal O}}
\def\cS{{\cal{S}}}
\def\cV{{\cal{V}}}
\def\cW{{\cal{W}}}

\def\e{{\mathfrak{e}}}

\def\ra{\rightarrow}
\def\Ra{\Rightarrow}

\def\d{\delta}
\def\a{{\alpha}}

\def\eps{\epsilon}

\def\E{\text{E}_{10}}
\def\K{K({\E})}

\def\lae{{\mathfrak{e}}}
\def\lah{{\mathfrak{h}}}
\def\lag{{\mathfrak{g}}}

\def\nn{\nonumber}

\def\cR{{\cal R}}
\def\cF{{\cal F}}
\def\cV{{\cal V}}

\def\cP{{\cal P}}

\def\pa{\partial}

\def\bx{{\bf{x}}}
\def\by{{\bf{y}}}

\newcommand{\eprint}[1]{{\href{http://arxiv.org/abs/#1}{[\texttt{#1}]}}}
\newcommand{\eprintN}[1]{{\href{http://arxiv.org/abs/#1}{[\texttt{#1 [hep-th]}]}}}
\newcommand{\eprintNT}[1]{{\href{http://arxiv.org/abs/#1}{[\texttt{#1 [math.NT]}]}}}
\newcommand{\eprintRT}[1]{{\href{http://arxiv.org/abs/#1}{[\texttt{#1 [math.RT]}]}}}
\newcommand{\eprintGR}[1]{{\href{http://arxiv.org/abs/#1}{[\texttt{#1 [gr-qc]}]}}}
\newcommand{\doi}[2]{{\hypersetup{urlcolor=darkred}\href{http://dx.doi.org/#2}{#1}\hypersetup{urlcolor=blue}}}

\begin{document}

\mbox{}
\vspace{10mm}

\begin{center}
{\bf \Large The $\E$ Wheeler--DeWitt operator at low levels}\\[12mm]
  Axel Kleinschmidt${}^{1,2}$ and Hermann Nicolai${}^1$\\[3mm]
\footnotemark[1]{\sl  Max-Planck-Institut f\"ur Gravitationsphysik\\
     Albert-Einstein-Institut \\
     M\"uhlenberg 1, D-14476 Potsdam, Germany\\[1mm]
\footnotemark[2]{\sl International Solvay Institutes\\
ULB-Campus Plaine CP231, BE-1050 Brussels, Belgium}\\[1mm]
E-mails: {\tt axel.kleinschmidt@aei.mpg.de}, {\tt nicolai@aei.mpg.de}} \\[27mm]
\begin{minipage}{12cm}\footnotesize
\textbf{Abstract:} 
We consider the Wheeler--DeWitt operator associated with the bosonic part of the
Hamiltonian of $D=11$ supergravity in a formulation with only the
spatial components of the three-form and six-form fields, and compare it
with the $\E$ Casimir operator at low levels, to show that these two operators 
precisely match modulo spatial gradients up to and including  $\mathfrak{gl}_{10}$ level $\ell =2$. 
The uniqueness of the $\E$ Casimir operator eliminates all ordering ambiguities 
in the quantum Hamiltonian, at least up to the level considered. 
Beyond $\ell\geq 3$ the two operators are expected to start to differ from each other, 
as they do so for the classical expressions.  We then consider 
truncations of the $\E$ Wheeler--DeWitt operator for various finite-dimensional 
subgroups of $\E$ in order to exhibit the automorphic properties of the
associated wave functions and to show that physically sensible 
wave functions generically vanish at the cosmological singularity, thus providing 
new and more sophisticated examples of DeWitt's proposed mechanism 
for singularity resolution in quantum gravity.  Our construction provides novel 
perspectives on several unresolved conceptual issues with the Wheeler--DeWitt 
equation, such as the question of observables in quantum gravity, or the 
issue of emergent space and time in a purely algebraic framework. We also 
highlight remaining open questions of the $\E$ framework.
\end{minipage}
\end{center}

\thispagestyle{empty}

\newpage
\setcounter{page}{1}

\tableofcontents

\section{Introduction}

The Wheeler--DeWitt (WDW) equation \cite{DeWitt,Wheeler} is the central equation 
of quantum  gravity (see \cite{Kiefer:2007ria} for an introduction and many 
further references). However, notwithstanding the fact that this equation has been 
around for more than half a century, only comparatively little progress has been 
achieved with it. This is due to conceptual issues (in particular concerning
the proper interpretation of the `wave function of the universe') as well as to 
severe mathematical difficulties that have so far thwarted all attempts to properly
formulate this equation in a manageable, mathematically well-defined 
and physically meaningful way.
To be sure, one can consider simplified versions of the WDW equation, 
such as the mini-superspace approximation often invoked in
quantum cosmology (see \cite{Kiefer:2007ria} for examples), or purely
topological theories, such as pure gravity in three space-time dimensions
(see {\em e.g.} \cite{Carlip}), or otherwise exploit the equation for heuristic purposes. 
However, none of these
simplifications addresses the core technical issue, namely the occurence of
short distance singularities which can be viewed as a non-perturbative
manifestation of the non-renormalisability of perturbative quantum gravity.
Furthermore, for pure gravity no observables in the sense of Dirac are known, 
that is, quantities commuting with all constraints, including the Hamiltonian constraint,
even though these are expected to be a central ingredient in any quantum 
mechanical setting. In addition, there is the crucial question as to
what extent space-time concepts are essential at the most fundamental 
level: indeed, many approaches to quantum gravity posit that space and time
should be emergent, rather than fundamental concepts, in which case field 
theoretic notions would lose all meaning at the Planck scale.

In this paper we take a new look at these issues and are motivated by 
unification and the view that a consistent theory of quantum gravity
requires the inclusion of very specific matter interactions determined by
symmetry considerations. More precisely, our approach is based
on the proposal of  \cite{Damour:2002cu} (see also \cite{Cocoyoc,KN0} for
reviews), according to which the maximal rank hyperbolic Kac--Moody 
algebra $\e_{10}$ should play a key role.\footnote{The hyperbolic Kac-Moody algebra
 relevant for pure (Einstein) gravity with zero cosmological constant in four space-time dimensions is the Feingold--Frenkel algebra  $AE_3$,
 while higher rank hyperbolic algebras appear for higher-dimensional (super-)gravities.
 However, out of these, only $\lae_{10}$ possesses a root lattice with  the self-duality 
 properties that we ultimately  expect to be required for the quantum consistency
 of the theory.}
This proposal builds on an old conjecture
\cite{Ju85} according to which this symmetry should appear as an extension
of the Geroch group in the reduction of maximal supergravity~\cite{Cremmer:1978km}
to one dimension. Although the present realisation takes place in a rather 
different and, in particular, {\em quantum mechanical} context we will see that 
the one-dimensional reduction fits very well with the WDW approach.

The standard canonical formulation of the bosonic sector of $D=11$ 
supergravity~\cite{Cremmer:1978km} is built on the spatial metric and the spatial three-form 
potential as well as their conjugate momenta as functions of a spatial coordinate 
${\bf x}$ \cite{Diaz:1986jw,Diaz:1986jx}.
By contrast, the $\E$ formulation has an infinity of fields, but no spatial dependence, because space is hypothesised to be an emergent concept in this approach~\cite{Damour:2002cu}.
In order to reconcile these two aspects, we strive to reformulate $D=11$ supergravity in a way that brings in new fields in addition to the usual ones, the first instance being a six-form potential dual to the three-form potential. Remarkably, there is a formulation, due to~\cite{Bunster:2011qp}, that uses 
only the spatial components of these two fields rather than a covariant three-form 
(see also~\cite{Bandos:1997gd} for related work). This formulation breaks manifest Lorentz invariance, but because this symmetry is also broken in the $\E$ approach,
such a formulation is much better suited for comparison with the $\E$ model
than the standard canonical formulation (which we also review here). 
More poignantly, we shall argue that this breaking of manifest Lorentz invariance 
is a {\em necessary prerequisite} for quantisation, as this is what allows 
us to treat the three- and six-form fields as independent degrees of freedom. 
Being off-shell and not manifestly Lorentz covariant, our approach differs from covariant ones such as the 
proposal of~\cite{West:2001as,Glennon:2020uov} for covariant 
and E$_{11}$-invariant equations of motions.

We will use this reformulation  to bring the bosonic Hamiltonian 
constraint of $D=11$ supergravity closer to an ultra-local form, by which we mean that only field values at a point enter but not their derivatives. This is done by considering the quantum analogue of the Belinsky--Khalatnikov--Lifshitz (BKL) limit~\cite{Belinsky:1970ew} where now in the operator realisation of the Hamiltonian constraint spatial derivatives of fields become negligible compared to functional derivatives rather than to time derivatives (as there is no pre-defined notion of time in canonical quantum gravity). As we shall explain in more detail in section~\ref{sec:quant}, this is related to the way the conjugate momenta (time derivatives) of the fields, including their duals, appear in the Hamiltonian constraint. 
After dropping the spatial gradients, the Hamiltonian then is ultra-local and depends only on canonical variables 
evaluated at one given spatial point ${\bf x}$ and no longer contains any spatial 
derivatives that connect neighbouring spatial points.
This form makes it possible to relate to the $\E$ context. The proper reinstatement of spatial gradients remains, however, to be clarified. We expect the other kinematic constraints (diffeomorphism and Gauss-type) appearing in field theory to play a crucial role in this.
However, their final significance in our present approach is 
not clear and might necessitate the introduction of extra fields beyond 
the $\E/\K$ symmetric space, in the same way that they appear in 
exceptional field theory~\cite{Hohm:2013jma,Hohm:2013pua,BKS}. 
A further indication of the need for extra degrees of freedom
is the apparent incompatibility of the full $\E/\K$ model
with supersymmetry \cite{Damour:2006xu,Kleinschmidt:2014uwa}.
The parts of the Gauss-type constraints that do not contain explicit spatial derivatives have been investigated in~\cite{DKN,Damour:2009ww} where it was found that they have a resemblance to Sugawara-type constructions, a fact that is also compatible with the extra fields of exceptional field theory and tensor hierarchy algebras~\cite{Bossard:2017wxl,BKS,Cederwall:2021ymp}. 

The Hamiltonian constraint, to the level checked in this paper and after 
dropping the spatial gradients, then coincides precisely 
with the invariant norm, {\em alias} the quadratic Casimir operator, 
appearing in the $\E$ model.   By the very definition
of the Casimir invariant, this means that we have access to an infinite number
of observables, at least in principle!
We expect our analysis to be extendable to the linearised dual graviton, 
keeping in mind the well known difficulties associated with this field~\cite{West:2001as,Hull:2001iu,Damour:2002cu,Bergshoeff:2008vc,Kleinschmidt:2014uwa}.   
However, as is 
evident from our derivation in section~\ref{sec:11D}, a complete analysis of the $\ell=3$ sector is
technically even more demanding than the present analysis, because various 
Dirac brackets between gravitational and matter variables will then no longer vanish.

Using the relation between the Hamiltonian constraint and the $\E$ approach we then proceed to propose a quantisation of $D=11$ supergravity that preserves this relation, {\em i.e.}, is based on maintaining $\E$ symmetry. This hinges on a realisation of the Lie algebra $\mathfrak{e}_{10}$ in terms of differential operators acting on the symmetric space $\E/\K$. Due to the infinite-dimensionality of the symmetric space and the presence of 
imaginary roots, the construction of such differential operators is difficult. For this reason
we shall content ourselves with a truncated version where we only 
consider the fields up to the six-form. 
Identifying the WDW-operator with the Casimir operator of $\mathfrak{e}_{10}$ in this truncation then provides an unambiguous quantisation of the system.
A further novel aspect is that, with the restriction to one spatial point, the
standard short distance singularities that usually hamper a proper definition 
of the WDW operator have entirely disappeared;  instead, one
has to cope with an exponentially growing tower of new degrees of freedom.
This situation is reminiscent of the one encountered in string theory: there as well, 
UV singularities are completely absent, but at each order in string perturbation
theory one must invoke a `division' by a modular group to render physical
quantities finite (besides ensuring absence of tachyons).
As we shall explain, there are hints of a similar mechanism at work in the present 
construction, which we shall make more explicit by studying certain truncations 
of the $\E$ WDW operator to a finite number of degrees of freedom
and by exhibiting the automorphic properties of the associated
wave functions (see \cite{Fleig:2015vky} for
an introduction to automorphic representation theory with a comprehensive
list of references). Such finite-dimensional truncations are instructive,
but it is not clear to what extent they can capture the full complexity 
of the $\E$ model because,  for the full theory, the relevant `modular 
group' is expected to be something vastly larger than the modular groups 
so far considered in string perturbation theory. Besides there are
new subtleties for imaginary roots, to be discussed in section~\ref{sec:imroot},
that cast doubt on the applicability of standard automorphic theory.
The use of a discretised version 
$\E(\mathbb{Z})$ of $\E(\mathbb{R})$, based on considerations
of BPS states, was already suggested in~\cite{Hull:1994ys}, and analysed further 
in relation to M-theory in~\cite{Ganor:1999ui,Brown:2004jb} with a viewpoint 
somewhat more similar to the one adopted here. Automorphic forms related 
to $\E$ have been studied in relation to scattering amplitudes in~\cite{Fleig:2012xa}.

Given the above reformulation of $D=11$ supergravity,
the central object of interest is the wave function $\Phi$ 
which is a function on the $\E/\K$ coset space.
A key question, even independently of the question whether extra fields beyond the coset ones 
are needed or not,  concerns the parametrisation of this wave function. 
We label the fields appearing in $\Phi$ according to \textit{all} elements that appear in the level decomposition of
$\lae_{10}$ with respect to its $\mathfrak{gl}_{10}$ subalgebra, that was also heavily used in 
 \cite{Damour:2002cu,Cocoyoc,KN0}. We note that there are different and inequivalent definitions of
Kac--Moody {\em groups} available, see~\cite{Kumar:2002,Marquis:2018} for overviews. The original proposal, now often called the minimal group, consisted of gluing SL(2,$\mathbb{R}$) subgroups of $\E$ associated to real roots only~\cite{KacPeterson}. In this parametrisation, a group element would consist of `words' made
 out of an infinite alphabet of real roots, together with appropriate relations inherited from the Serre relations in the Lie algebra~\cite{Kumar:2002,Marquis:2018}. The parametrisation used in physics~\cite{Damour:2002cu,Cocoyoc,KN0} associates one field component to each positive root generator, for both real \textit{and} imaginary roots. This picture is much closer to the maximal Kac--Moody group defined in the literature~\cite{Marquis:2018}. Which global parametrisation and which choice of group $\E$ is most suitable to physics, and, 
more specifically, which choice is best suited for explaining the emergence of spatial
dependence,  is not clear at the moment and deserves further study. In this paper 
we work in level decomposition. 

One of the main outcomes of our analysis is a more detailed understanding
of the asymptotic behaviour of the wave function as it approaches what would be a 
(space-like) classical singularity. Previous work on this~\cite{Kleinschmidt:2009cv,Kleinschmidt:2009hv} based on a quantisation of only the cosmological billiard (BKL) approximation of the classical singularity revealed that the wave function vanishes in this limit, thus realising DeWitt's idea of a quantum-mechanical resolution of classical singularities.\footnote{We note that this statement, as well as the chaoticity of the classical cosmological billiard rely on the structure of $D=11$ supergravity which in particular excludes a cosmological constant. Similar statements hold 
for pure bosonic $D=4$ gravity with zero cosmological constant. Therefore our results are 
not in conflict with the no-boundary proposal of~\cite{Hartle:1983ai}
which requires a non-zero cosmological constant.}
 We shall show in section~\ref{sec:qBKL} that this 
result is robust when including more degrees of freedom compared to BKL. 
Even though our analysis is still in a truncated setting we take this as an indication 
that the full quantum $\E$ system (with appropriate discrete symmetries) could 
be a sensible model of quantum gravity.
The behaviour of the wave function near the singularity may also have  
implications for the information paradox. In \cite{Perry:2021mch} it is argued 
that information cannot be lost if it is not crushed in the singularity (although 
it remains unclear how it would be released again upon Hawking evaporation).
However, even if the wave function in a BKL type approximation as
in (\ref{BKL}) vanishes at the singularity,  there remains the question as to its
behavior when infinitely many degrees of freedom `open up' at the singularity \cite{Nic}.
The latter possibility is strongly suggested by the fact that 
{\em classical} geodesics on the $\E/\K$ manifold are infinitely 
unstable along directions involving imaginary root spaces \cite{DN},
although it is an open question whether and how this instability percolates 
to the quantised theory.

There remain two major unsolved problems at this point. The first is
to understand the emergence of spatial dependence beyond the use of 
dual fields, and thus to extend the `dictionary' of \cite{Damour:2002cu}
beyond first order spatial gradients.
Namely, like all previous ones, our calculation 
systematically ignores spatial gradients other than those obtained via dualisation. 
A significant observation in this context may be our equation (\ref{A3A6}) which 
states that the Dirac bracket between the three- and six-form fields only vanishes
up to spatial gradients (besides being non-local). Discarding the latter is thus 
consistent with our level expansion (\ref{cV}) and (\ref{cN}) for {\em commuting} 
fields. However, in the quantised theory these two fields can no longer
be treated as commuting ($c$-number) objects, because
\beq\label{A3A6quant}
\big[ \hat{A}_{m_1m_2m_3}(\bx)\,,\, \hat{A}_{n_1\cdots n_6}(\by) \big] \,=\,  - 
i\hbar \GN \, \ve_{m_1m_2m_3 n_1\cdots n_6 p}  \pa^p G(\bx,\by)  
\eeq
where $G(\bx,\by)$ is the scalar propagator. This result, which ties the appearance 
of spatial gradients to an emergent non-commutativity of the basic variables,
may indicate the need for some kind of non-commutative geometry on 
the $\E/\K$ manifold, as well as for third quantisation (in the sense that the
wave function $\Phi$ in (\ref{Psi}) may become operator valued).
We also notice that the right-hand side of (\ref{A3A6quant}) brings 
in both $\hbar$ and the Newton constant $\GN$, and thus a notion of length 
which is not present in the dimensionless pre-geometrical setting.

The second open problem concerns the proper incorporation of fermions 
into the $\E/\K$ model, in a way that is fully compatible with $\E$ symmetry 
(or an even larger framework), and that can also capture spatial 
dependence (there are no known analogs of dual fields
for fermions). First steps in this direction were already taken some time
ago, by showing that the gravitino components at a fixed spatial point make 
up a finite-dimensional unfaithful spinorial representation of the R-symmetry 
group $\K$~\cite{deBuyl:2005zy,Damour:2005zs,deBuyl:2005sch,Damour:2006xu},
and by re-interpreting the $D=11$ Rarita--Schwinger equation as a
$\K$ covariant `Dirac equation' \cite{Damour:2006xu}. Fermions have 
also been included in mini-superspace approaches to $N=1$ supergravity 
in $D=4$ in~\cite{Damour:2013eua,Damour:2014cba} and $D=5$ \cite{Damour:2022oja}
where they were found to be compatible with singularity avoidance. 
Treating fermions in the standard way would modify the WDW equation by 
fermionic contributions, and thus spoil the identification of the WDW operator 
with the $\E$ Casimir operator already at the very lowest order (as is evident
from the explicit expressions given in~\cite{Damour:2013eua,Damour:2014cba}).
Moreover, even neglecting spatial dependence, it would blow up the 
scalar wave function $\Phi$ 
to an object with $2^{160}$ 
components~\cite{Kleinschmidt:2009cv,Kleinschmidt:2009hv}. 
While fermions are not immediately necessary for arriving at the $\E$ conjecture, 
it seems clear that their inclusion will be essential for the consistency of the
full theory and for singling out better quantum behaviour, much in the same way
that local supersymmetry improves the behaviour of perturbative quantum supergravity, 
and that fermions are needed in string theory for finiteness via 
modular invariance and elimination of tachyons. For the present model, the main 
question is therefore whether it is possible to include fermions in a way that 
manifestly preserves the $\E$ structure of the WDW Hamiltonian; this may require
a novel type of bosonisation, perhaps along the lines of \cite{Nahm}.
In any case, we expect the answer to these questions to also have
a bearing on other outstanding issues. 

The structure of this paper is as follows. We first analyse the bosonic sector of $D=11$ supergravity canonically, recasting the matter sector in a `democratic' form, where both a three-form and a six-form appear. The canonical quantisation of the resulting theory is then studied and the Wheeler--DeWitt equation worked out. In section~\ref{sec:E10}, we then consider the functional realisation of $\E$ in terms of differential operators. The formal Casimir operator is then worked out in this language and compared to the WDW operator of $D=11$ supergravity in section~\ref{sec:comp}. We relate our results on full (super-)gravity to previous work on its cosmological billiards truncation in section~\ref{sec:qBKL}, where we also highlight the different effects of real and imaginary roots on solutions and connections to theory of automorphic forms. Section~\ref{sec:cmts} contains concluding general remarks on properties of the wave function.

Throughout this paper we employ units with $c=1$. For further
reference let us record the dimensions of the various objects. We have 
$[\hbar]=M L$ (mass$\times$length); with dimensionless fields and coordinates of dimensions $L$ (length), the $D=11$ Newton constant has dimension 
$[\GN]=L^{8}M^{-1}$, so the Planck length is $\ell_{\rm P}= (\hbar \GN)^{1/9}$. The conjugate momenta have dimension $[\Pi]=M L^{-9}$. The delta density has dimension $[\delta(\bx)]=L^{-10}$ which is also the dimension of the functional derivative $[\delta/ \delta \phi(\bx)] =L^{-10}$ for any dimensionless field $\phi$.

\section{\texorpdfstring{Bosonic Hamiltonian of $D\!=\!11$ supergravity}{Bosonic Hamiltonian of D=11supergravity}}
\label{sec:11D}

In this section, we analyse the bosonic sector of $D=11$ supergravity~\cite{Cremmer:1978km} by first performing the standard Hamiltonian treatment of the metric and the three-form field (see for instance~\cite{Diaz:1986jw}). To bring the resulting expressions closer to the $\E$ model, we then introduce a dual six-form potential and reformulate
the canonical theory in  a version where only the {\em spatial} components
of the three- and six-form fields are retained, following~\cite{Bunster:2011qp}.\footnote{This formulation is also sometimes called the Henneaux--Teitelboim form because of~\cite{Henneaux:1988gg,Henneaux:1987hz}.} 
In principle this reformulation can also be applied to the gravitational sector,
but we leave this step to future work.
The resulting expressions are then quantised canonically as a further preparation for comparison with a functional realisation of $\E$.

\subsection{\texorpdfstring{Bunster--Henneaux form of $D=11$ supergravity}{Bunster--Henneaux form of D=11 supergravity}}

We start from the bosonic part of the $D=11$ supergravity Lagrangian
in the conventions of~\cite{Damour:2006xu}
\begin{align}
\label{eq:SUGRA11}
\GN \mathcal{L} & \,=\,  \frac{E}4   R \,-\, \frac{E}{48} F^{MNPQ} F_{MNPQ}
+ \, \frac{2}{(144)^2} \ve^{M_1\ldots M_{11}} F_{M_1\ldots M_4} F_{M_5\ldots M_8} A_{M_9M_{10}M_{11}}
\end{align}
with $F_{MNPQ} \equiv 4 \pa_{[M} A_{NPQ]}$ and $E$ the determinant of the elfbein $E_M{}^A$. These fields depend on
eleven coordinates $x^M \equiv (t, {\bf x})$ (with time $t$, and where ${\bf x}$ 
coordinatises the spatial hypersurface) which we usually do not write out. We have explicitly written an overall factor of the Newton constant $\GN$ 
to emphasise that all bosonic field are dimensionless, as required for the
comparison with the $\E/\K$ sigma model where the coset degrees of freedom
are likewise dimensionless.
In (\ref{eq:SUGRA11}), $\ve^{M_1\ldots M_{11}}$ is the numerical Levi--Civita \textit{symbol}, a space-time tensor \textit{density} for which we use the convention that $\ve^{0\,1\ldots 10}= - \ve_{0\,1\dots 10} = +1$. 
The Levi--Civita \textit{tensor} with upper and lower indices given by 
$\epsilon^{M_1\ldots M_{11}} = E^{-1} \ve^{M_1\ldots M_{11}}$
and $\epsilon_{M_1\ldots M_{11}} = E \ve_{M_1\ldots M_{11}}$, with 
analogous definitions for the purely spatial objects.

For the elfbein we assume the triangular gauge
\beq
\label{eq:elfbein}
E_M{}^A \,=\, 
\begin{pmatrix}
N \,&\, e_m{}^a N^m\\
0 \,&\,  e_m{}^a
\end{pmatrix}
\qquad \Ra \qquad E = Ne \quad\quad 
\eeq
with the lapse function $N$ and the shift $N^m$, which
are, respectively, the Lagrange multipliers associated to the Hamiltonian and
diffeomorphism constraints; $e=\det e_m{}^a$ is the volume density of 
the ten-dimensional spatial slice. 
We split curved space-time indices as $M=(t,m)$ and flat ones as $A=(0,a)$. The Levi--Civita symbol on a spatial slice is induced from that in space-time by $\ve^{m_1\ldots m_{10}}\equiv \ve^{tm_1\ldots m_{10}}$. A further constraint will be seen to be the Gauss constraint associated to the 
Lagrange multiplier field $A_{tmn}$.

For the canonical treatment we first determine the canonical momenta conjugate to
$g_{mn}$ and $A_{mnp}$,
respectively, which are given by\footnote{The conjugate momenta of $N$, $N^m$ and $A_{tmn}$ vanish as primary constraints. As is usual for $p$-forms coupled to gravity, the fields serve as Lagrange multipliers for the first-class constraints that generate the corresponding gauge transformations.}
\begin{align}
\label{eq:CM11}
\GN \, \Pi^{mn} & =\frac12   e e^{am} e^{bn} \big( \Omega_{0(ab)} - \delta_{ab} \Omega_{0cc} \big)
\,,\nn\\
\GN \, \Pi^{mnp} &=  - E F^{tmnp} 
+\frac1{216}\ve^{tmnpk_1\ldots k_7}F_{k_1\ldots k_4} A_{k_5k_6k_7}
\end{align}
where $\Omega_{ABC} \equiv E_A{}^M E_B{}^N(\pa_M E_{NC} - \pa_N E_{MC})$ are the $D=11$
coefficients of anholonomy and have only flat indices (see {\em e.g.}  \cite{MN} for further explanations and
conventions as well as~\cite{Henneaux:1986cz,Diaz:1986jw,Diaz:1986jx,Kreutzer:2020quf} for canonical treatments of supergravity theories). For obtaining the above relations we have used the functional derivatives in the normalisations
\begin{align}
\label{eq:fnder}
\frac{\delta g_{mn}({\bf x})}{\delta g_{pq}({\bf y})} = 2 \delta^p_{(m} \delta^q_{n)}  \delta({\bf x},{\bf y})\,,\quad\quad
\frac{\delta A_{m_1\ldots m_p}({\bf x})}{\delta A_{n_1\ldots n_p}({\bf y})} = 
p!\,  \delta^{n_1\, \ldots\, n_p}_{m_1\ldots m_p} \delta({\bf x},{\bf y})
\end{align}
with $\delta^{n_1\, \ldots\, n_p}_{m_1\ldots m_p} = \delta^{[n_1}_{[m_1}\cdots \delta^{n_p]}_{m_p]}$ and (anti-)symmetrisations of unit strength. The 
spatial Dirac deltas $\delta(\bx,\by)$ appearing on the right-hand sides are densities with respect 
to spatial diffeomorphisms and satisfy $\int d^{10} \by f(\by) \delta(\bx,\by) = f(\bx)$.
The momenta defined in~\eqref{eq:CM11} are also tensor \textit{densities},  and likewise for the functional derivative operator $\frac{\delta}{\delta \phi(x)}$ for all fields $\phi(x)$.
We also note the appearance of the Newton constant $\GN$ in the relation between
the velocities and  the canonical momenta. 
From the way the Newton constant appears in the definition of the canonical momenta~\eqref{eq:CM11}, we see that it is important for relating the canonical momenta to spatial derivatives of the fields.

 For notational convenience we also define the part of the three-form 
canonical momentum due to the Chern--Simons term by
\begin{align}
\mathcal{P}^{mnp} = \frac1{216}\ve^{mnpk_1\ldots k_7}F_{k_1\ldots k_4} A_{k_5k_6k_7}\,,
\end{align}
where the ten-dimensional Levi--Civita symbol is related to the eleven-dimensional one by $\ve^{m_1\ldots m_{10}} \equiv \ve^{tm_1\ldots m_{10}}$.
With this definition we have
\begin{align}
F^{tmnp} = - E^{-1} \left(\GN \Pi^{mnp} -\cP^{mnp}\right)\,.
\end{align}
The standard canonical treatment then leads to the Hamiltonian form of the action
\begin{align}
\label{eq:can11}
\mathcal{L}_{\text{can}} \,=\, \frac12 \dot{g}_{mn} \Pi^{mn}  + \frac1{3!} \dot{A}_{mnp} \Pi^{mnp} - N \mathcal{H} -N^m \mathcal{H}_m - \frac12 A_{tmn} \mathcal{G}^{mn} \,,
\end{align}
The (first-class) constraints appearing in~\eqref{eq:can11} are  
\begin{align}
\label{eq:H11}
e \mathcal{H}&\,=\,  \GN  G_{mn|pq} \Pi^{mn} \Pi^{pq} - \frac1{4\GN} e^2 R^{(10)} \nn\\[2mm]
& \qquad + \frac{1}{12\GN} \big(\GN \Pi^{mnp} - \mathcal{P}^{mnp} \big) g_{mm'}g_{nn'}g_{pp'}
 \big(\GN \Pi^{m'n'p'} - \mathcal{P}^{m'n'p'} \big) \nn\\[2mm]
&\qquad +\frac{1}{48\GN}e^2 F_{m_1\ldots m_4}g^{m_1n_1}\cdots g^{m_4n_4}F_{n_1\ldots n_4}\,,
\end{align}
and
\begin{subequations}
\label{eq:C11}
\begin{align}
\label{eq:diff}
\mathcal{H}_m &= - g_{mn }\nabla_p \Pi^{pn} + \frac16 F_{mnpq}( \Pi^{mnp}-\GN^{-1} \mathcal{P}^{mnp})\,,\\[2mm]
\label{eq:Gauss}
\mathcal{G}^{mn} &=  -\partial_p \Pi^{pmn} -\frac{\GN^{-1}}{12\cdot 144} \ve^{mnk_1\ldots k_8} F_{k_1\ldots k_4}F_{k_5\ldots k_8}\nn\\
&= \partial_p \left[ - \Pi^{pmn} -\frac{\GN^{-1}}{3\cdot 144} \ve^{mnpk_1\ldots k_7} A_{k_1k_2k_3} F_{k_4\ldots k_7}\right]\,,
\end{align}
\end{subequations}
where $\nabla_p$ is the covariant derivative with the Levi--Civita connection for the spatial metric $g_{mn}$
(note that the gradient in (\ref{eq:Gauss}) is an {\em ordinary} derivative because
$\Pi^{mnp}$ transforms as a density).
They are, in turn, the Hamiltonian (scalar) constraint, the diffeomorphism (momentum) constraint 
and the Gauss constraint. These constraints must be imposed at each point $\bf{x}$ of the spatial
hypersurface, whence we are dealing with a continuous infinitude of constraints.
Observe that we have pulled out a factor of $e$,  
as a result of which the `potential terms' appear with a prefactor $e^2$. 
The DeWitt metric for $D=11$ space-time dimensions is (here
without a density factor)
\begin{align}
\label{eq:DWM}
 G_{mn|pq} \,: = \, g_{p(m}g_{n)q}   - \frac19  g_{mn} g_{pq}\,.
\end{align}

As is well-known, the equations of motion for the matter field allow the introduction of a dual six-form potential $A_{M_1\ldots M_6}$ on-shell. While it is not possible to write the non-linear theory covariantly in terms of 
only the six-form potential~\cite{Nicolai:1980kb} (or even both covariant potentials at the same time), it was shown in~\cite{Henneaux:1988gg,Bunster:2011qp} that one can write an off-shell theory without manifest Lorentz and diffeomorphism covariance when using only the spatial components $A_{m_1m_2m_3}$ and $A_{m_1\ldots m_6}$ of both potentials\footnote{For a closely related formulation with an extra vector
allowing for a general `axial' gauge see \cite{Bandos:1997gd}.};
the remaining manifest symmetry is the SO(10)
subgroup of the Lorentz group. We review this formalism in some detail 
in appendix~\ref{app:HT} where we follow~\cite{Bunster:2011qp}. Such a formulation 
is desirable because these are exactly the fields that appear in the $\E$ theory which 
does not exhibit manifest Lorentz symmetry either.  Although 
the notion of level will be `officially' introduced only in section~\ref{sec:GL10dec}, 
we already here refer to the metric, three-form, six-form and dual graviton
fields as ``level-$\ell$ fields", for resp. $\ell=0,1,2,3$, see also (\ref{EEE}).

As derived in appendix~\ref{app:HT}, the action of~\cite{Bunster:2011qp} involving 
both spatial potentials in our conventions is
\begin{align}
\label{eq:HT11}
\mathcal{L}_{\text{can}} &= \frac12 \dot{g}_{mn} \Pi^{mn}  + \frac{\GN^{-1}}{2\cdot 3!\cdot 7!} \dot{A}_{mnp}\ve^{mnpk_1\ldots k_7} F_{k_1\ldots k_7} - \frac{\GN^{-1}}{2\cdot 4! \cdot 6!} F_{mnpq} \ve^{mnpqk_1\ldots k_6} \dot{A}_{k_1\ldots k_6}
\nn\\
&\quad 
+ \frac{\GN^{-1}}{3!\cdot 864} \dot{A}_{mnp} \ve^{mnpk_1\ldots k_7} A_{k_1k_2k_3} F_{k_4\ldots k_7}
 - N \mathcal{H} -N^m \mathcal{H}_m \,,
\end{align}
where the terms with time derivatives on the three- and six-form result from the
replacement of $\Pi^{mnp}$ by the solution (\ref{eq:GaussSol}) of the Gauss constraint~\eqref{eq:Gauss} which has therefore disappeared. The Hamiltonian 
constraint is now given by
\begin{align}
\label{eq:H36}
e \mathcal{H} &\,= \GN \,  G_{mn|pq} \Pi^{mn} \Pi^{pq} -\frac{\GN^{-1}} 4 e^2 R^{(10)} +\frac{\GN^{-1}}{2\cdot 4!} e^2 F_{m_1\ldots m_4}g^{m_1n_1}\cdots g^{m_4n_4}F_{n_1\ldots n_4}\nn\\
&\quad 
+ \frac{\GN^{-1}}{2\cdot 7!}e^2 F_{m_1\ldots m_7} g^{m_1n_1}\cdots g^{m_7n_7} F_{n_1\ldots n_7} \,,
\end{align}
and where 
\begin{align}
\label{eq:F7def}
F_{m_1\ldots m_7} = 7\partial_{[m_1} A_{m_2\ldots m_7]}  -35 A_{[m_1m_2m_3} F_{m_4\ldots m_7]}
\end{align}
is the spatial field strength of the six-form potential, including a coupling to the 
three-form due to the Chern--Simons term, cf. (\ref{eq:defF7}).
To arrive at this form, the original Gauss constraint~\eqref{eq:Gauss} has been solved and this is the step that introduces the spatial six-form $A_{m_1\ldots m_6}$.
In the form~\eqref{eq:H36}, the Hamiltonian still depends on the magnetic field strengths $F_{m_1\ldots m_4}$ and $F_{m_1\ldots m_7}$ that contain in particular spatial derivatives. As the $\E$ theory does not directly contain spatial derivatives, we now  manipulate the theory further. We note that in the Bunster--Henneaux form~\eqref{eq:HT11} 
no temporal components of the gauge fields appear explicitly.   To
simplify the subsequent canonical analysis of the matter sector {\em we now switch to a
flat background geometry in the remainder of this subsection}, anticipating that for 
the final results we can re-convert to general backgrounds by covariantising 
the relevant expressions. For this reason we will also set $\GN=1$ until
the end of this subsection, but re-instate $\GN$ in the following sections.

The theory~\eqref{eq:HT11} contains the primary constraints  
\begin{subequations}
\label{C0}
\begin{align}
\label{eq:C03}
\cC^{mnp}(\bx) \,&\coloneqq\,  \Pi^{mnp} -
\frac1{2\cdot 7!} \ve^{mnpk_1\ldots k_7} F_{k_1\ldots k_7} - \frac1{3!\cdot 144} \ve^{mnpk_1\ldots k_7} A_{k_1k_2k_3} F_{k_4\ldots k_7}\,,
 \\[2mm]
\label{eq:C06}
\cC^{m_1\cdots m_6} (\bx) \,&\coloneqq \, 
\Pi^{m_1\cdots m_6}  +  \frac1{2\cdot 4!} \ve^{m_1\ldots m_6 k_1\ldots k_4} F_{k_1\ldots k_4}\,,
\end{align}
\end{subequations}
where now the momenta are defined from the Lagrangian density~\eqref{eq:HT11}.

For the further analysis we need the non-vanishing Poisson brackets of the elementary variables 
which are normalised as
\begin{align}
\{ A_{mnp}(\bx), \Pi^{qrs}(\by) \}_{\rm PB} &= 3!\, \d^{qrs}_{mnp} \d(\bx,\by)\,,\nn\\[2mm]
\{ A_{m_1\cdots m_6}(\bx), \Pi^{n_1\cdots n_6}(\by) \}_{\rm PB} &= 
6!\, \d^{n_1\cdots n_6}_{m_1\cdots m_6} \d(\bx,\by).
\end{align} 
The Poisson bracket itself carries a dimension of inverse action, i.e., $[\{\cdot,\cdot\}]= M^{-1}L^{-1}$. The same will be true for the Dirac bracket to be written below.

With this one can work out the matrix of Poisson brackets of the constraints~\eqref{C0} as
\begin{align}
\label{CC}
&\quad \begin{pmatrix}
\big\{ \cC^{m_1m_2 m_3} (\bx)\,,\, \cC^{n_1n_2 n_3} (\by)\big\}_{\rm PB}  &
\big\{ \cC^{m_1m_2 m_3}(\bx) \,,\, \cC^{n_1\cdots n_6} (\by)\big\}_{\rm PB} 
\\[2mm]
\big\{ \cC^{m_1\cdots m_6}(\bx)  \,,\, \cC^{n_1n_2 n_3} (\by)\big\}_{\rm PB}\; &
\;  \big\{ \cC^{m_1\cdots m_6} (\bx) \,,\, \cC^{n_1\cdots n_6} (\by) \big\}_{\rm PB} 
\end{pmatrix} 
\\[3mm]
&  = \; \begin{pmatrix}
\frac1{24} \, \ve^{m_1m_2m_3n_1n_2n_3 p_1\cdots p_4} F_{p_1\cdots p_4} (\bx) 
 \delta(\bx,\by) 
& -\ve^{m_1m_2m_3n_1\cdots n_6 p} \pa_p \delta(\bx,\by) \\[3mm]
-  \ve^{m_1 \cdots m_6n_1n_2 n_3 p} \pa_p \delta(\bx,\by)  & 0 
\end{pmatrix}    \nn\,,
\end{align}
where by convention the derivative on the $\d$-function always acts on the
first argument (the antisymmetry of this matrix under simultaneous interchange
of indices and coordinates follows from $\pa_\bx \d(\bx,\by) = - \pa_\by \d(\by,\bx)$).

Demanding that the constraints~\eqref{C0} be preserved in time leads to conditions on the 
associated Lagrange multipliers introduced for them in the canonical formalism. 
The Lagrange multipliers are then fixed up to homogeneous solutions that are independent of the canonical variables. Hence, there are no further (secondary) constraints to be considered.

The following Hamiltonian density generates the same matter dynamics as the Lagrangian~\eqref{eq:HT11} when taken in flat space
\begin{align}
\label{eq:HamPB}
\cH &= \frac{1}{2\cdot 4!} F_{n_1\ldots n_4} F^{n_1\ldots n_4} + \frac{1}{2\cdot 7!} F_{n_1\ldots n_7} F^{n_1\ldots n_7} + \frac1{3!\cdot 7!} \ve_{n_1n_2n_3m_1\ldots m_7} \cC^{n_1n_2n_3} F^{m_1\ldots m_7} \nn\\
&\quad + \frac1{6!} \cC^{n_1\ldots n_6} \bigg[ -\frac1{4!} \ve_{n_1\ldots n_6m_1\ldots m_4} F^{m_1\ldots m_4} + \frac1{144}  \ve_{n_1\ldots n_6m_1\ldots m_4} F^{m_1\ldots m_7} A_{m_5m_6m_7} \bigg] \nn\\
&\quad + \frac12 \lambda_{mn} \cG^{mn} + \frac1{5!} \lambda_{n_1\ldots n_5} \cG^{n_1\ldots n_5}\,.
\end{align}
We emphasise that the dynamics are generated using Poisson brackets and that they 
are \textit{weakly} equal to the Euler--Lagrange equations deduced from~\eqref{eq:HT11}. 
In particular, the field dependent coefficients of $\cC^{\cdots}$ in (\ref{eq:HamPB}) are chosen in such a way that $\dot \cC^{\cdots} \approx0$.
The free gauge parameters $\lambda_{mn}$ and $\lambda_{n_1\ldots n_5}$ can also be viewed as being related to the temporal components of the three- and six-form fields.

As our goal is to quantise the system, we need to follow the Dirac formalism and work with Dirac rather than Poisson brackets. The transition to Dirac brackets in particular removes part of phase space and therefore reduces the number of variables, making the resulting expression closer to the $\E$ approach.
The constraints~\eqref{C0} represent a mixed system of first- and second-class constraints.  
This can be deduced from the fact that the matrix~\eqref{CC} is degenerate, as one can see by acting 
with it on the vector\footnote{\label{fn:comp}This action contains both a contraction of the tensorial indices with the canonical combinatorial factors $1/3!$ and $1/6!$ as well as an integral over $\by$ as shown in~\eqref{eq:GG}.}
\beq\label{Nullmod}
\Big(v_{n_1n_2n_3} \,\Big|\, v_{n_1\ldots n_6}\Big)^T =  \Big(3 \p_{[n_1} \lambda_{n_2n_3]}(\by)\, \Big| \, 
6\p_{[n_1} \lambda_{n_2\cdots n_6]}(\by) +15 \,
  F_{[n_1n_2n_3n_4} (\by) \lambda_{n_5n_6]}(\by)  \Big)^T\;.
\eeq
This null vector corresponds to first-class constraints $\cG$ according to
\begin{align}
\label{eq:GG}
\cG &= \int \!\!d \by \left[ \frac1{3!} v_{n_1n_2n_3} \cC^{n_1n_2n_3} + \frac1{6!} v_{n_1\ldots n_6} \cC^{n_1\ldots n_6}\right]
=   \int\!\! d \by \left[ \frac1{2} \lambda_{mn} \cG^{mn} + \frac1{5!} \lambda_{m_1\ldots m_5} \cG^{m_1\ldots m_5}\right]
\end{align}
for any $\lambda_{m_1m_2}(\by)$ and $\lambda_{m_1\ldots m_5}(\by)$ of dimension $L$. The integrated generator $\cG$ is only non-trivial for $\lambda_{m_1m_2}$ and $\lambda_{m_1\ldots m_5}$ that are non-trivial in de Rham cohomology, i.e., that cannot be written as $\lambda_{m_1m_2} = 2\partial_{[m_1} \sigma_{m_2]}$ in terms of a one-form $\sigma_m$ (and similarly a four-form $\sigma_{m_1\ldots m_4}$ for $\lambda_{m_1\ldots m_5}$). This well-known gauge-for-gauge structure is equivalent to the reducibility of the gauge constraints. Our analysis will not depend on resolving this reducible structure. 

The local generators of gauge transformations associated with~\eqref{eq:GG} can be read off as
\begin{subequations}
\label{eq:G2G5}
\begin{align}
 \cG^{mn}(\bx) & \,\coloneqq\, - \pa_p \cC^{mnp}(\bx) +\frac1{24} \cC^{mnp_1\ldots p_4} (\bx)F_{p_1\ldots p_4}(\bx)\,,\\[2mm]
\cG^{m_1\cdots m_5} (\bx) &\,\coloneqq \,  - \,  \pa_p \cC^{pm_1\cdots m_5} (\bx) \,.
\end{align}
\end{subequations}
In a similar way, one can write a projection to the second-class constraints, that we denote by $\cS^{\cdots}$, in the form 
\begin{subequations}
\label{eq:tC0}
\begin{align}
\label{eq:tC03}
\cS^{n_1n_2n_3} (\bf x)\,&\coloneqq \,\cC^{n_1n_2n_3} (\bx)
-3 \int d\by \, \partial^{[n_1} G(\bx,\by) \, \partial_p  \cC^{n_2n_3]p}(\by) 
\nn\\
&\quad + A_{p_1\ldots p_3} (\bx) \int d\by \, \partial^{[n_1} G(\bx,\by) \, \partial_k \cC^{n_2n_3p_1\ldots p_3] k} (\by)\nn\\
&\quad 
-\frac12 \int d\by \, \partial^{[n_1}G(\bx,\by)\,  \partial_k \left( \cC^{n_2n_3] p_1\ldots p_3 k} A_{p_1\ldots p_3}(\by)\right)\\
\label{eq:tC06}
\cS^{n_1\ldots n_6}(\bx) 
\,&\coloneqq \,\cC^{n_1\ldots n_6}(\bx) +   6  \int d\by \, \partial^{[n_1} G(\bx,\by) \, \partial_p \cC^{n_2\ldots n_6]p}(\by)
\end{align}
\end{subequations}
where $G(\bx,\by)$ is the spatial (flat) Green function satisfying $\Delta G(\bx,\by)  = \partial_m\partial^m G(\bx,\by) = \delta(\bx,\by)$.\footnote{In Cartesian coordinates on ten flat Euclidean dimensions (with vanishing conditions at infinity), the Green function can be written explicitly as $G(\bx,\by) = (\textrm{vol}(S^9))^{-1}  |\bx-\by|^{-8}$, but we shall not rely on this expression. The properties of it that we use are its defining Laplace equation together with 
$G(\bx,\by)=G(\by,\bx)$ and $\partial_\bx G(\bx,\by) = - \partial_\by G(\bx,\by)$. }
All derivatives act on the first argument of a function unless indicated otherwise. 
The dimension of the Green function is $L^{-8}$. The appearance
of the scalar propagator in these expressions already points to one extra complication 
with {\em curved} backgrounds: because of the implicit dependence of $G(\bx,\by)$ 
on the spatial metric the gravitational momenta $\Pi^{mn}$ then no longer commute 
with the Green function.

The true second-class generators~\eqref{eq:tC0} therefore contain non-local terms. 
Using the convolution product defined by $(f\star g)(\bx) \equiv \int d\by f(\bx,\by) g(\by)$, 
 we can thus separate the original constraints (\ref{eq:C03})
and (\ref{eq:C06}) into first- and second-class constrains as 
\begin{subequations}
\label{eq:splitC}
\begin{align}
\cC^{n_1n_2n_3} & \,=\, \cS^{n_1n_2n_3} - 3\,  \partial^{[n_1}G \star \cG^{n_2n_3]} -\, A_{p_1p_2p_3} \left( \partial^{[n_1} G \star \cG^{n_2n_3 p_1p_2p_3]}\right)\nn\\
&\quad +\frac12 \partial^{[n_1}G \star \left(\cG^{n_2n_3]p_1p_2p_3}  A_{p_1p_2p_3}  \right)
\,,\\
\cC^{n_1\ldots n_6} &\,=\,  \cS^{n_1\ldots n_6} -  6 \,\partial^{[n_1} G \star \cG^{n_2\ldots n_6]}\,,
\end{align}
\end{subequations}
where we have also expressed the non-local terms through the Gauss constraints~\eqref{eq:G2G5} in order to exhibit that the second-class constraints differ from the full constraints by terms proportional to the first-class constraints.

We note some properties of the first- and second-class constraint that are important for the further development of the Dirac formalism.
First, and by construction of the null vector, the first-class generators $\cG^{\cdots}$ Poisson commute with all constraints~\eqref{C0} in the strong sense, {\em viz.}
\bea
\label{eq:FCcom}
\big\{ \cC^{mnp} (\bx) \,,\, \cG^{q_1q_2}(\by) \big\}_{\rm PB} \,&=&\, 
\big\{ \cC^{mnp} (\bx)\,,\, \cG^{q_1\cdots q_5}(\by) \big\}_{\rm PB} \,=\, 0 \,, \nn\\[2mm]
\big\{ \cC^{m_1\cdots m_5}(\bx) \,,\, \cG^{q_1q_2} (\by)\big\}_{\rm PB} \,&=&\, 
\big\{ \cC^{m_1\cdots m_5}(\bx) \,,\, \cG^{q_1\cdots q_5}(\by) \big\}_{\rm PB} \,=\, 0\,.
\eea
In particular, the first-class constraints all (strongly) commute with one another.
We also note that the second Gauss-type constraint $\cG^{m_1\ldots m_5}$ is automatically 
divergence-free while $\cG^{mn}$ satisfies
\begin{align}
\partial_m \cG^{mn} \,=\,  \frac1{24} \partial_m \cC^{mnp_1\ldots p_4} F_{p_1\ldots p_4}  \,=\,
 - \frac1{24} \cG^{np_1\cdots p_4} F_{p_1\cdots p_4}
\,.
\end{align}
The gauge transformations generated by the first-class combinations (\ref{eq:G2G5}) are 
\begin{align}
\label{eq:GT}
\delta A_{n_1n_2n_3} = 3\partial_{[n_1} \lambda_{n_2n_3]} \,,
\hspace{10mm}
\delta A_{n_1\ldots n_6} =6 \partial_{[n_1} \lambda_{n_2\ldots n_6]} + 15 F_{[n_1\ldots n_4} \lambda_{n_5n_6]}
\end{align}
and one can check that they leave the field strengths $F_{n_1\ldots n_4}$ and $F_{n_1\ldots n_7}$ (defined in~\eqref{eq:F7def}) appearing  in the Hamiltonian~\eqref{eq:H36} invariant. The appearance of the gauge-invariant field strength $F_{m_1\ldots m_4}$ means that in this basis the gauge algebra is abelian. This corresponds to the fact that the first-class constraints $\cG^{\cdots}$ strongly Poisson commute with all constraints and therefore also $\{ \cG^{n_1n_2}, \cG^{n_3n_4}\}_{\rm PB}=0$ strongly.\footnote{Alternatively, one could also choose a basis where this commutator only vanishes weakly modulo $\cG^{n_1\ldots n_5}$ by taking $\cG^{n_1n_2}\to \cG^{n_1n_2}+ (*) A_{m_1m_2m_3} \cG^{n_1n_2m_1m_2m_3}$, leading to a non-abelian gauge algebra.}

The second-class constraints~\eqref{eq:tC0} satisfy the relations
\begin{subequations}
\label{eq:SCfn}
\begin{align}
- \partial_p \cS^{pmn} + \frac1{24} \cS^{mn p_1\ldots p_4} F_{p_1\ldots p_4} &= 0\,,\\
-\partial_p \cS^{pm_1\ldots m_5} &=0\,,
\end{align}
\end{subequations}
so that inserting them into the projection~\eqref{eq:G2G5} to the first-class constraints gives zero. In fact, this property is what was used to determine the expressions~\eqref{eq:tC0}.
Since the second-class constraints differ from the full constraints $\cC^{\cdots}$ by first-class constraints and the first-class constraints strongly Poisson commute with all constraints according to~\eqref{eq:FCcom}, we deduce that the matrix of Poisson brackets of the second-class constraints is identical to~\eqref{CC}. However, it is now to be thought of as acting on the space of second-class functions, {\em i.e.}, those satisfying~\eqref{eq:SCfn}.

To determine the Dirac brackets, we shall now invert the matrix~\eqref{CC} 
on this function space. We repeat that we work
on flat space-time in order to render the expressions more tractable, 
as  the following discussion 
only concerns the gauge structure in the matter sector. With the full gravitational
couplings, the analysis of the second-class constraints and the determination of
the Dirac brackets become substantially more complicated due to their 
dependence on the spatial metric and the resultant non-commutativity
with the gravitational momenta. Nevertheless, we shall see that the
relevant expressions can be covariantised in a straight-forward manner and
thus extended to curved space, but we leave a detailed analysis of the associated 
subtleties to future work.

The inverse of the matrix (\ref{CC}) on the space of second-class functions is given by
\begin{align}
\label{eq:invCS}
\begin{pmatrix}
     0 &
  -\ve_{m_1m_2m_3 n_1\cdots n_6 p} \pa^p G (\bx,\by) \\[3mm]
 - \ve_{m_1 \cdots m_6 n_1n_2 n_3 p} \pa^p G(\bx,\by)  & 
  X_{m_1\cdots m_6 \, n_1\cdots n_6} (\bx,\by)
\end{pmatrix} \,,
\end{align}
where 
\begin{align}
X_{m_1\ldots m_6 \, n_1\cdots n_6} (\bx,\by) \,&=\,
-20\, \ve_{km_1\ldots m_6 [n_1n_2n_3} A_{n_4n_5n_6]}(\by) \partial^k G(\bx,\by)\nn\\
&\quad\quad 
-20\, \ve_{kn_1\ldots n_6 [m_1m_2m_3} A_{m_4m_5m_6]}(\bx) \partial^k G(\bx,\by)
\end{align}
satisfies $X_{m_1\ldots m_6 \, n_1\ldots n_6} (\bx,\by) = - X_{n_1\ldots n_6 \, m_1\ldots m_6} (\by,\bx)$.\footnote{The matrix operator~\eqref{eq:invCS} is the inverse of~\eqref{CC} on the space of second-class constraints, i.e., its composition with~\eqref{CC} in the sense of footnote~\ref{fn:comp} acts as the identity on functions $(\cS^{n_1n_2n_3},  \cS^{n_1\ldots n_6})$ satisfying the relation~\eqref{eq:SCfn}. It is \textit{not} the inverse on the space of all functions $(\cC^{n_1n_2n_3},\cC^{n_1\ldots n_6})$ where an inverse does not exist due to the degeneracy of~\eqref{CC}. Therefore the product of the matrices~\eqref{CC} and~\eqref{eq:invCS} is not the identity matrix but it acts as the identity on the relevant space.}

With these preparations one can now work out the Dirac brackets. These become quite long and they also contain non-local terms that are due to the separation into first- and second-class constraints we have chosen in~\eqref{eq:splitC}. We shall only give a few salient ones for the elementary variables and then focus on the gauge-invariant objects whose Dirac brackets are free of these non-local terms~\cite{Henneaux:1987hz}.
We have for instance the following Dirac brackets among position and momentum variables
\begin{align}
\label{Dirac}
\big\{ A_{mnp}(\bx)\,,\, \Pi^{qrs} (\by)\big\}_{\rm DB} \, &= \,
\frac12 \cdot 3! \,\Big(  \d_{mnp}^{qrs} \d(\bx,\by)
+ 3 \, \d^{[qr}_{[mn} \pa_{p]} \pa^{s]} G(\bx,\by) \Big)
\nn\\[2mm]
\big\{ A_{mnp}(\bx) \,,\, \Pi^{q_1\cdots q_6} (\by) \big\}_{\rm DB} \, &= \, 0 
\\[2mm]
\big\{ A_{m_1\cdots m_6}(\bx)\,,\, \Pi^{n_1\cdots n_6} (\by)\big\}_{\rm DB} \, &= \,  
\frac12 \cdot 6! \, \Big(  \d_{m_1\cdots m_6}^{n_1\cdots n_6}  \d(\bx,\by)
 \,+\,  6 \,\d^{[n_1\cdots n_5}_{[m_1 \cdots m_5} \pa_{m_6]} \pa^{n_6]} G(\bx,\by) \Big) \nn
 \end{align}
as well as non-trivial relations among only position and only momentum variables
 \begin{align}
 \label{A3A6}
\big\{ A_{m_1m_2m_3}(\bx)\,,\, A_{n_1\cdots n_6}(\by) \big\}_{\rm DB} \,&=\,  - 
\ve_{m_1m_2m_3 n_1\cdots n_6 p}  \pa^p G(\bx,\by)  \,,\\[2mm]
\big\{ \Pi^{m_1m_2m_3}(\bx)\,,\, \Pi^{n_1\cdots n_6}(\by) \big\}_{\rm DB} \,&=\, \ve^{m_1m_2m_3 n_1\cdots n_6 p}  \pa_p G(\bx,\by) \,.\nn
\end{align}
These Dirac brackets agree with those one would obtain in the free theory without the Chern--Simons coupling. The other brackets, such as $\big\{ A_{m_1\cdots m_6}(\bx)\,,\, \Pi^{qrs} (\by)\big\}_{\rm DB}$, are non-vanishing only for non-zero Chern--Simons coupling. As an example we have
\begin{align}
\big\{ \Pi^{m_1m_2m_3} (\bx) ,\,\Pi^{n_1n_2n_3} (\by) \big\}_{\rm DB} &=
 \frac1{4\cdot 6!\cdot 6!} \ve^{m_1m_2m_3p_1\ldots p_7} \ve^{n_1n_2n_3s_1\ldots s_7} \partial_{p_7}^x \partial_{s_7}^y X_{p_1\ldots p_6\, s_1\ldots s_6}(\bx,\by)\,.
\end{align}

Turning to the Dirac brackets involving the gauge-invariant field strengths $F_{m_1\ldots m_4}=4\partial_{[m_1} A_{m_2m_3m_4]}$ and $F_{m_1\ldots m_7}$ defined in~\eqref{eq:F7def}, we find
\begin{align}\label{AF}
\big\{ A_{m_1 m_2 m_3}(\bx)\,,\, F_{n_1\cdots n_7}(\by) \big\}_{\rm DB} \,&=\,
  +  \, \ve_{m_1 m_2 m_3 n_1\cdots n_7} \d(\bx,\by)  
    \,  -3 \, \ve_{p n_1\cdots n_7[m_1m_2}  \pa_{m_3]} \pa^p G(\bx,\by)  
  \\
\big\{ A_{m_1 \cdots m_6}(\bx)\,,\, F_{n_1\cdots n_4}(\by) \big\}_{\rm DB} \,&=\,
   \,  - \, \ve_{m_1 \cdots m_6 n_1\cdots n_4} \d(\bx,\by)  
   \,  - 6 \,  \ve_{p n_1\cdots n_4 [m_1\cdots m_5}  \pa_{m_6]} \pa^p G(\bx,\by) \,.\nn
\end{align}
The Dirac bracket $\{A_{m_1\ldots m_6}(\bx), \, F_{n_1\ldots n_7}(\bx) \}_{\rm DB}$ is also non-zero in the interacting theory with Chern--Simons term. However, most important
for our analysis are the Dirac brackets among the gauge-invariant field strengths 
which take a simpler form, {\em viz.}
\begin{subequations}
\label{eq:FFDB}
\begin{align}\label{FF1}
\big\{ F_{m_1\cdots m_4}(\bx)\,,\, F_{n_1\cdots n_4}(\by) \big\}_{\rm DB} \,&=\, 0  \,, \nn\\[2mm]
\big\{ F_{m_1\cdots m_4}(\bx)\,,\, F_{n_1\cdots n_7}(\by) \big\}_{\rm DB} \,&=\,  -
7 \, \ve_{m_1\cdots m_4 [n_1 \cdots n_6} \pa_{n_7]}\d(\bx,\by) 
\end{align}
and
\begin{align}\label{FF2}
\big\{ F_{m_1\cdots m_7}(\bx)\,,\, F_{n_1\cdots n_7}(\by) \big\}_{\rm DB} \,=\,  
- \frac1{432} \ve_{m_1\cdots m_7 p_1p_2p_3} \ve_{n_1\cdots n_7 p_4p_5p_6} 
\ve^{p_1\cdots p_6 q_1\cdots q_4} F_{q_1\cdots q_4}(\bx) \d(\bx,\by)\,.
\end{align}
\end{subequations}
In particular they are local, in the sense that they contain only $\d$-functions
or derivatives of $\d$-functions, but no Green functions. Furthermore, 
these brackets are straightforward to convert into generally covariant form 
 since the right-hand sides of the brackets do not 
depend on the metric and are tensorial.

We can now rewrite the dynamics by using the Dirac brackets. Referring back to the Hamiltonian~\eqref{eq:HamPB}, we know that it generates the correct dynamics (weakly) when using Dirac brackets that also allow us to set to zero the second-class constraints in the Hamiltonian. Since the terms involving the constraints $\cC^{\cdots}$ in~\eqref{eq:HamPB} decompose according to~\eqref{eq:splitC} into first- and second-class pieces, we see that the first-class Lagrange multiplier terms in~\eqref{eq:HamPB} acquire additional field-dependent and non-local contributions. As these are, however, pure gauge transformations, we can absorb their effect in a redefinition of the Lagrange multipliers. Working out the dynamics in the Dirac formalism up to gauge transformations, we only need to consider the Hamiltonian density
\begin{align}
\label{eq:Hmm}
\cH_{\text{mat}} = \frac{1}{2\cdot 4!} F_{n_1\ldots n_4} F^{n_1\ldots n_4} + \frac1{2\cdot 7!} F_{n_1\ldots n_7}F^{n_1\ldots n_7}\,.
\end{align}
If one were to also substitute the second-class constraints into this expression, one would obtain terms involving $\Pi^2$ but also non-local terms, and for this reason we keep the expression above.

Given these Dirac brackets and the Hamiltonian (\ref{eq:Hmm}), we can now
check that the Hamiltonian equations of motion
\begin{align}
\label{EoM}
\dot A_{mnp}(\bx)  \,&\approx\, \int d\by \, \big\{ A_{mnp}(\bx), \, \cH_{\text{mat}}(\by) \big\}_{\rm DB} 
   \,+\, 3\, \pa_{[m} A_{np]t}(\bx) \nn\\[2mm]
\dot A_{m_1\cdots m_6}(\bx)  \,&\approx \, \int d\by \, \big\{  A_{m_1\cdots m_6}(\bx), \, \cH_{\text{mat}}(\by)  \big\}_{\rm DB} 
   \,-\, 6 \, \pa_{[m_1} A_{m_2\cdots m_6]t}(\bx) \nn\\[2mm]
   & \hspace{2cm} + \, 35 \, A_{[t m_1 m_2} F_{m_3\cdots m_6]}(\bx)
\end{align}
reproduce the duality relation (\ref{eq:dual11}).
Here, we have absorbed all gauge terms in the definition of the temporal component of the matter potentials.
Furthermore, the second relation is consistent with the result for $A_{tmn}$ from the first line.

Our analysis above was restricted to the matter sector and we worked in flat space 
for simplicity. For the comparison with $\E$ beyond level $\ell =2$ one will need to 
treat gravity in a similar way, by introducing the dual graviton field 
$A_{m_0|m_1\cdots m_8}$.
At the linearised level, it is possible to perform a similar replacement of the 
spatial derivatives of the metric by the momentum conjugate to the dual graviton~\cite{West:2001as,Hull:2001iu,Henneaux:2004jw,Julia:2005ze}.
As this extension involves substantial technical complications we 
leave the inclusion of this field to future work.

As a final remark we note that is tempting to substitute the \textit{full} constraints $\cC^{\cdots}=0$ from~\eqref{C0} into the Hamiltonian~\eqref{eq:HamPB}. This would lead directly to a form of the Hamiltonian that is quadratic in the momenta $\Pi$ and moreover ultra-local in that it only depends on canonical variables at one spatial point and is free of derivatives (except for the gauge transformations). This procedure, however, is inconsistent with the Dirac algorithm. 

\subsection{Canonical quantisation: a new perspective}
\label{sec:quant}

Given the Dirac brackets (\ref{Dirac}) and~\eqref{eq:FFDB} we can now 
proceed to canonical quantisation, replacing the canonical variables by 
functional derivative or multiplication operators in the standard way.
For the metric we have the usual substitutions
\beq
\hat{g}_{mn} (\bx) \,=\, g_{mn}(\bx) \;\;, \quad
\hat{\Pi}^{mn} (\bx) \,=\, i \hbar \frac{\delta}{\delta  g_{mn}({\bf x})} 
\eeq
where operators are indicated with hats.

For the three- and six-form fields the rules must be adapted in order to account
for the non-vanishing commutators (\ref{A3A6}) and similar non-vanishing Dirac brackets among the momenta. As these brackets depend on the split of the constraints into first- and second-class constraints and contain non-local term, we focus rather on the Dirac brackets~\eqref{eq:FFDB} among the gauge-invariant field strengths. For obtaining a quantisation of the Hamiltonian~\eqref{eq:Hmm}, this is sufficient. 
Operators that realise the algebra~\eqref{eq:FFDB} are  
\begin{align}
\label{eq:Fop}
\hat{F}_{m_1\ldots m_4}(\bx) &= -\frac{2}{6!} i\hbar \GN\,
\ve_{m_1\ldots m_4 n_1\ldots n_6} \frac{\delta}{\delta A_{n_1\ldots n_6}(\bx)}  
+\frac{10}3
\,\pa_{[m_1} A_{m_2 m_3 m_4]}(\bx)
\nn\\[2mm]
\hat{F}_{m_1\ldots m_7}(\bx) &= -  \frac{2}{3!} i\hbar \GN \,
\ve_{m_1\ldots m_7 n_1n_2n_3} \left( \frac{\delta}{\delta A_{n_1n_2n_3}(\bx)} +\frac1{12} A_{s_1s_2s_3}(\bx) \frac{\delta}{\delta A_{s_1s_2s_3n_1n_2n_3}(\bx)}\right) \nn\\[1mm]
&\hspace{15mm}-\frac73 \pa_{[m_1} A_{m_2\ldots m_7]}(\bx)
+\frac{140}{3} A_{[m_1m_2m_3} \pa_{m_4} A_{m_5m_6m_7]}(\bx)\,.
\end{align}
We see that these operator realisations involve a mix of functional derivatives and 
multiplicative operators. 
These expressions are tensorial and hold also on curved spaces. 
The Dirac brackets of the
non-gauge invariant fields and momenta can be similarly worked out, but contain
non-local contributions. We shall not give the explicit expressions here as they are not needed.

After replacement of the classical quantities by the above operators the quantum Hamiltonian can be presented in the simple operatorial form
\beq\label{eq:QH0}
\hat \cH(\bx) \,=\, \hat \cH_0(\bx)\, \,- \frac14 e R^{(10)}(\bx)
\eeq
where 
\begin{align}
\label{QH1}
e\hat\cH_0(\bx) \,&=\,  - \hbar^2 \GN G_{mn|pq} (\bx) \frac{\d}{\d g_{mn}(\bx)} 
\frac{\d}{\d g_{pq}(\bx)}\,  \nn \\
&\quad +\;
\frac{\GN^{-1}}{2\cdot 4!}  e^2 \hat{F}_{n_1\ldots n_4} (\bx) g^{n_1m_1}(\bx) \cdots g^{n_4m_4} (\bx) \hat{F}_{m_1\ldots m_4}(\bx) \\
&\quad 
+ \frac{\GN^{-1}}{2\cdot 7!} e^2 \hat{F}_{n_1\ldots n_7}(\bx) g^{n_1m_1}(\bx) \cdots g^{n_7m_7}(\bx)  \hat{F}_{m_1\ldots m_7}(\bx)\,. \nn
\end{align}
and where we have separated off the potential term involving the Ricci scalar
in (\ref{eq:QH0}) because at this point the gravitational
sector is described solely in terms of the metric. If a full non-linear dualisation of gravity were employed one would expect to also distribute the dynamics more democratically between the metric and its dual field. At the linearised level, the dual field to the $D=11$ metric has tensor structure $A_{m_0|m_1\ldots m_8}$~\cite{West:2001as,Hull:2001iu} and is well-known to appear in the $\mathfrak{gl}_{10}$ level decomposition of $\lae_{10}$~\cite{Damour:2002cu,Nicolai:2003fw} as we shall review in section~\ref{sec:E10}. 
The ordering of operators in (\ref{QH1}) is still arbitrary at this point.
We take it to be the one written and this will be seen to agree with the ordering 
which is uniquely fixed by the $\E$ Casimir operator in the next section.

The above differential operators are then supposed to act on the 
`wave function of the universe'  $\Psi$, which for the theory~\eqref{eq:HT11} 
in question is a {\em functional} of the variables $g_{mn}({\bf x})$,  
$A_{mnp}({\bf x})$ and $A_{m_1\cdots m_6}(\bx)$ (and eventually also
the dual graviton $A_{m_0|m_1\cdots m_8}(\bx)$). Upon making 
the requisite operator replacements in the Hamiltonian constraint 
and assuming that $\cH(\bx)$ can be properly defined as a quantum 
operator we end up with the WDW equation
\beq\label{WDW0}
\hat{\cH}({\bf x})\Psi \,=\, 0\;. 
\eeq
In addition, the wave functional $\Psi$ must satisfy the kinematic constraints 
\beq\label{kinConstr}
\hat\cH_m(\bx) \Psi \,=\, \hat\cG^{mn}(\bx) \Psi \,=\, 
\hat\cG^{m_1\cdots m_5} (\bx)\Psi \,=\, 0 \;.
\eeq 
These constraints must be imposed for all spatial points labelled by $\bx$.  
Extending the procedure of the foregoing section also to the gravitational sector 
would result in another constraint, supplementing the kinematic 
constraints (\ref{kinConstr}) by a `dual' spatial diffeomorphism constraint 
(corresponding to the Bianchi identity on the spatial curvature tensor). 

As we already mentioned there has been only scant progress with these 
equations due to conceptual and mathematical problems,  such as the 
`clash' of functional differential operators at coincident points in (\ref{QH1}). 
As we will  now see, the present reformulation offers an entirely new 
perspective on these problems. Namely,
we propose to replace the wave {\em functional} $\Psi$ above by a wave {\em function} 
$\Phi$ depending on infinitely many variables corresponding to the 
degrees of freedom in the coset space $\E/\K$, that is
\beq\label{Psi}
\Psi\Big[ g_{mn}({\bf x}) , A_{mnp}({\bf x}), \dots \Big] \;\;
 \longrightarrow \quad
 \Phi\big( g_{mn} , A_{mnp}, A_{mnpqrs}, A_{m_0|m_1\cdots m_8}, \ldots \big) 
\eeq
The main step is thus to replace a set of field variables depending on 
the spatial coordinates $\bx$  by an infinite tower of new variables corresponding 
to the degrees of freedom present in the $\E/\K$ coset space, and which depend 
no longer on $\bx$; the dots in the argument of $\Psi$ are included in order
to allow for further dual and auxiliary field variables (which, however,  
cannot change the on-shell content of the theory). At least in principle, 
the arguments in the new wave function $\Phi$ are supposed 
to correspond to the values of $g_{mn}({\bf x})$ and $A_{mnp}({\bf x})$, 
and their duals {\em at one fixed spatial point}  ${\bf x} = {\bf x}_0$, as
well as possibly other degrees of freedom. 
This identification is accompanied by the replacement
of functional differential operators by ordinary partial derivatives
according to the rule
\beq\label{FDer}
\frac{\d}{\d \phi(\bx_0)} \; \to \; \ell_{\rm P}^{-10} \frac{\pa}{\pa \phi}  \qquad\qquad
\mbox{for $\phi = g_{mn}, A_{mnp}, \cdots$}
\eeq
where $\ell_{\rm P} \equiv (\hbar\GN)^{1/9}$ is the (eleven-dimensional) Planck
length. Observe that this is {\em not} a discretisation in any standard 
sense as there is no underlying 
space lattice here: rather the spatial dependence is supposed to get
encoded into the infinite tower of dual variables on which $\Phi$ 
depends.\footnote{This is also suggested by the reduction
 of maximal supergravity to {\em two} space-time dimensions, where E$_9$ takes 
 the place of $\E$. There, at least at the level of the equations
 of motion, the coordinate dependence of the basic fields can 
 be encoded into an {\em infinite} tower of dual potentials, which in principle 
 allows us to extract the information on spatial dependence from the dependence 
 on the spectral parameter at a given spatial point (see \cite{NS} and references therein).}
This effective reduction to one spatial point is in accord with the $\E/\K$
sigma model proposal of~\cite{Damour:2002cu}, where the $D=11$ theory is reduced 
to one dimension, and the (first order) spatial gradients of the basic fields 
are regarded and treated as independent degrees of freedom.
In other words,  spatial dependence has been traded for an infinity of variables
at a fixed spatial point, but these can be associated directly with spatial gradients
only in lowest order of the level expansion \cite{Damour:2002cu,Cocoyoc}.

Neglecting spatial gradients is usually associated
with the BKL limit in the classical theory, where the Einstein equations are 
supposed to be dominated by time derivatives near the cosmological 
singularity \cite{Belinsky:1970ew}.
When extending such considerations to the quantised theory, 
it is important to keep in mind that in the WDW approach there is 
no {\em a priori} `time', unlike for a Schr\"odinger wave function, and 
hence also no hidden time dependence in any of the canonical expressions. 
Rather, time is supposed to emerge operationally by picking a `clock variable' 
and by approximating the WDW equation in a semi-classical expansion by an 
effective Schr\"odinger equation, as for instance explained in \cite{Kiefer:2007ria}.
With regard to (\ref{eq:Fop}), we therefore take the BKL limit as being  equivalent 
to neglecting spatial gradients in comparison with the functional differential 
operators, since the latter originate from canonical momenta, which themselves
are related to velocities. This limit does not involve the Planck length.

The expression that will be related to $\E$ in the next section is then~\eqref{eq:Fop} with the spatial gradients dropped. Before going into the technical details, we sketch what the correspondence will be.
As a first step, and keeping in mind
the caveats mentioned above and in the introduction, we propose that 
the standard WDW equation (\ref{WDW0}) 
should be replaced by a new constraint equation
\beq\label{WDW1}
\Omega \, \Phi \,=\, 0
\eeq
where the original Hamiltonian $\hat{\cH}(\bx)$ is replaced by the $\E$ Casimir operator
$\Omega$, up to overall factors, see~\eqref{H=Om} below. This operator, whose 
differential operator realisation will  be discussed in detail in the 
following sections, acts on the $\E/\K$ coset  space degrees of freedom 
which appear as arguments of $\Phi$.  Importantly, in this version of the theory 
any reference to `space' has disappeared! Likewise there are no short distance 
singularities any more, thanks to the replacement of functional derivative
operators by ordinary derivatives with respect to the coset variables, cf. (\ref{FDer}).
Another key feature 
is that  the $\E$ Casimir operator is {\em unique} \cite{Kac:1968,Kac}, and 
therefore {\em the proper  operator ordering is pre-ordained by $\E$ symmetry}.\footnote{In 
contrast to finite-dimensional Lie algebras, $\E$ admits no polynomial Casimir 
operators other than the quadratic one \cite{Kac}.} Our main result in~\eqref{H=Om} below then is that  
the operators $\hat{\cH}_0(\bx_0)$ for fixed $\bx_0$
and $\Omega$ match precisely up to and including 
level $\ell=2$. This non-trivial agreement extends partially to level $\ell=3$ 
if one replaces the spin connection by  the `dual graviton' variable 
$A_{m_0|m_1\cdots m_8}$ (where however the trace $\omega_{bba}$ of the spin connection is 
missing due to the constraint $A_{[m_0|m_1\cdots m_8]}=0$)
and the  spatial curvature term in $\cH$ by yet another kinetic term.  In the same way that the six-form arose from solving locally the Gauss constraint~\eqref{eq:Gauss}, the linearised dual graviton will arise from solving locally the linearised diffeomorphism constraint~\eqref{eq:diff}.

Of course, many questions remain, even disregarding the issue of fermions. 
One is the fact that the matching between $\cH_0$ and $\Omega$ 
fails starting from level $\ell =3$, reflecting the incompleteness of the `dictionary' 
presented in \cite{Damour:2002cu}. This incompleteness is also evident from 
the fact that the spatial Ricci scalar can assume both positive and negative 
values, whereas the $\E$ Casimir is a positive operator away from the 
Cartan subalgebra. Furthermore, it seems doubtful that the discrepancies 
arising at levels $|\ell|\geq 3$ can be resolved purely within the framework
of $\E$ alone, as already suggested by the absence of the trace of the spin connection.
It has been argued from the point of view of exceptional field theory and the tensor hierarchy algebra that an appropriate extension of the $\E$ coset will involve an indecomposable structure where $\E$ is augmented by highest weight representations where the first one is triggered by the trace of the spin connection~\cite{Bossard:2017wxl,BKS,Cederwall:2021ymp}, a fact that is also suggested by compatibility with supersymmetry~\cite{Damour:2006xu,Bossard:2019ksx}.

The present approach thus suggests 
that the notion of `space' must be  extracted in a similarly `operational' way
as the notion of `time'. For this we would need to incorporate the kinematical
constraints also into the $\E$ framework by endowing them with a group
theoretical realisation. First steps in this direction were
taken in \cite{DKN} where an attempt was made to assign
these constraints to a representation of $\E$ (which however cannot be a highest
or lowest weight representation). If this could be done, we would re-interpret the
group theoretical version of the diffeomorphism constraint operator $\cH_m$
as a generator of spatial coordinate dependence, simply by conjugation with
the operator $\exp(\xi^m \cH_m)$\footnote{It has been 
observed that the gauge parameters $(\xi^m, \xi_{mn},
\xi_{m_1\cdots m_5},  \cdots)$ associated
to the kinematic constraints (\ref{kinConstr}) constitute  the beginning of 
the $\Lambda_1$ representation of $\E$~\cite{West:2003fc,Damour:2007dt,Coimbra:2011ky,Berman:2012vc,GGN}},
where $\xi^m$ is some coordinate parametrising the motion away from 
$\bx_0 \equiv \bx(0)$.  Although such formulas are familiar from quantum 
field theory,  the crucial difference is that the operator $\cH_m$ would here be 
defined entirely group theoretically, and without reference to
a pre-existing space-time structure, unlike the standard momentum operator
in quantum field theory.

We end this section by observing that in the full theory, the dimensionful constants $\GN$ and $\hbar$ appear explicitly in the operator realisations~\eqref{eq:Fop} in between the functional derivatives and multiplication parts of the operators. As we shall in the next section consider the operators at one fixed spatial point and drop all spatial derivatives, the operators become homogeneous in the dimensionful constant that thus can be eliminated from the WDW equation~\eqref{WDW0}. This is in agreement with the fact that the $\E$ model does not contain any dimensionful constants.

\section{\texorpdfstring{Functional realisation of $\E$ at low levels}{Functional realisation of E10 at low levels}}
\label{sec:E10}

In this section, we first explain some basic features related to the 
level expansion of the hyperbolic Kac--Moody algebra $\e_{10}$, see~\cite{Kac,KacMoodyWakimoto,Damour:2002cu,Nicolai:2003fw,Cocoyoc} 
for more information. Then we proceed to realise the beginnings of this algebra 
formally in terms of differential operators on an  infinite-dimensional function space 
by considering its action on the symmetric space $\E/K(\E)$. Of course,
this is still very far from providing a proper understanding of $\lae_{10}$: 
continuing with the construction one quickly runs into the very same difficulties
as with more traditional realisations, because the full differential operators
are unmanageable infinite sums whose summands contain an exponentially
growing number of terms..

\subsection{\texorpdfstring{$\E$ commutation relations at levels $|\ell|\leq 3$}{E10 commutation relations at levels <=3}}
\label{sec:GL10dec}

There is no known explicit representation of the $\E$ Lie algebra (this
remains the key unsolved problem in the theory of indefinite Kac--Moody 
algebras since its inception more than 50 years ago \cite{Moody:1968,Kac:1968,Kac}). 
Some insight can be gained by decomposing it in terms of representations
of a `manageable' subalgebra. This is achieved by making a
level decomposition, which is a $\mathbb{Z}$-graded decomposition 
of the infinite-dimensional Lie algebra 
\beq\label{level0}
\lae_{10} = \bigoplus_{\ell= -\infty}^\infty \lae^{(\ell)}_{10}
\eeq
There are different choices for the $\ell =0$ subalgebra, but here we pick
the one best adapted to the problem at hand, namely
\beq
\lae^{(0)}_{10} \,=\, \mathfrak{gl}_{10}
\eeq
see~\cite{Damour:2002cu,Nicolai:2003fw,Cocoyoc} for further details and 
explanations; we mostly follow notation and conventions of \cite{Cocoyoc}.
Other possible choices for the level-0 subalgebra  $\lae_{10}^{(0)}$ are 
$\mathfrak{so}(9,9)\oplus \mathfrak{gl}_1$ and $\mathfrak{gl}_9 \oplus \mathfrak{sl}_2$ and, 
respectively, correspond to type IIA and type IIB supergravity \cite{KN1,KN2}.

The $\mathfrak{gl}_{10}$ generators $K^m{}_n$ obey the standard commutation 
relations
\beq
\big[ K^m{}_n\,,\, K^p{}_q \big] \,=\, \d_n^p K^m{}_q - \d_q^m K^p{}_n
\eeq
The associated standard bilinear form $\langle \cdot | \cdot \rangle$ reads
\beq
\langle  K^m{}_n | K^p{}_q \rangle = \d^m_q \d^p_n - \d^m_n \d^p_q\,,
\eeq
where the trace term proportional to $\d^m_n\d^p_q$ is left
undetermined by $\mathfrak{gl}_{10}$ and gets fixed 
only after embedding $\mathfrak{gl}_{10}\subset\lae_{10}$.
At levels $\ell=1,2,3$ the subspaces $\lae_{10}^{(\ell)}$ are, respectively,
spanned by a three-form, a six-form and mixed Young tableau representation, to wit,
\beq\label{EEE}
E^{m_1m_2m_3} \;,\, E^{m_1\cdots m_6} \;,\, E^{m_0|m_1 \cdots m_8} 
\eeq
The corresponding negative level generators for $\ell = -1,-2,-3$ are
\beq\label{FFF}
F_{m_1m_2m_3} \;,\, F_{m_1\cdots m_6} \;,\, F_{m_0|m_1 \cdots m_8} 
\eeq
with $E^{[m|n_1\cdots n_8]} = F_{[m|n_1\cdots n_8]} = 0$. We note that
these representations are in one-to-one correspondence with the
ones encountered in the Hamiltonian analysis of the foregoing section.
Yet higher level generators in (\ref{level0})
can be determined analogously, but the analysis becomes rapidly 
more  complicated, and actually unmanageable beyond the very lowest levels,
see {\em e.g.} \cite{Nicolai:2003fw} for
a table of representations up to $|\ell|\leq 28$.\footnote{Corresponding tables
for $\lae_{10}^{(0)} = \mathfrak{so}(9,9)\oplus\mathfrak{gl}_1$ can be found in \cite{KN1} and for the type IIB case $\lae_{10}^{(0)} = \mathfrak{gl}_9\oplus \mathfrak{sl}_2$ in~\cite{KN2}.}
For $|\ell|\leq 3$ the commutation relations between positive and negative level 
generators read
\bea
\big[ E^{mnp} , E^{qrs} \big]  &=& E^{mnpqrs} \;\; , \quad
\big[ E^{mnp} , E^{q_1\cdots q_6} \big] = 3 E^{[m|np]q_1\cdots q_6}   \nn\\[2mm]
\big[ F_{mnp} , F_{qrs} \big]  &=& - F_{mnpqrs} \;\; , \quad
\big[ F_{mnp} , F_{q_1\cdots q_6} \big] = - 3 F_{[m|np]q_1\cdots q_6}
\eea
Note that the normalisation of the $\ell=3$ generator differs from \cite{Cocoyoc}
by a factor of 3. The minus sign in the definition of $F_{mnpqrs}$ and $F_{m|n_1\cdots n_8}$ ensures that (formally)
the $E$'s and $F$'s are each other's hermitean conjugates: $E^\dagger = F$.
These generators transform in the standard tensorial way under GL(10):
\beq
\big[ K^m{}_n \,,\, E^{qrs} \big] \,=\, 3 \,\d^{[q}_n E^{rs]m} \;\; , \quad
\big[ K^m{}_n \,,\, F_{qrs} \big] \,=\, - 3 \, \d_{[q}^m F_{rs]n} \;,  \quad \text{etc.}
\eeq
For the commutators mixing positive and negative levels we have
\beq
\big[ F_{mnp} \,,\, E^{qrs} \big] \,=\, -18 \,\d^{[qr}_{[mn} K^{s]}{}_{p]}
                           + 2 \, \d^{qrs}_{mnp} K\,,
\eeq
where $K\equiv K^m{}_m$, and
\bea
\big[ F_{m_1m_2 m_3}\,,\, E^{n_1\cdots n_6} \big]\, &=&\,
5! \, \d_{m_1m_2 m_3}^{[n_1n_2n_3} E^{n_4n_5n_6]}         \nn\\[2mm]
\big[ F_{m_1\cdots m_6}\,,\, E^{n_1n_2 n_3} \big]\, &=&\,
5! \, \d^{n_1n_2 n_3}_{[m_1m_2m_3} F_{m_4m_5m_6]}         \nn\\[2mm]
\big[ F_{m_1\cdots  m_6}\,,\, E^{n_1\cdots n_6} \big]\, &=&\,
- 6\cdot 6!  \, \d_{[m_1 \cdots m_5}^{[n_1\cdots n_5} K^{n_6]}{}_{m_6]} 
+ \frac23 \cdot 6!\, \d^{n_1\cdots n_6}_{m_1\cdots m_6} K \nn\\[2mm]
\big[ F_{m_1m_2 m_3}\,,\, E^{n_0|n_1\cdots n_8} \big]\, &=&\,
7\cdot 16 \, \Big(\d^{n_0|[n_1n_2}_{m_1m_2m_3} E^{n_3\cdots n_8]}  \,-\,
\d^{[n_1n_2n_3}_{m_1m_2m_3} E^{n_4\cdots n_8]n_0}  \Big)
\nn\\[2mm]
\big[ F_{m_1\cdots m_6}\,,\, E^{n_0|n_1\cdots n_8} \big]\, &=&\,  \frac13\cdot 8!\,
\Big( \d_{m_1\cdots m_6}^{n_0[n_1\cdots n_5} E^{n_6n_7n_8]} \,-\, 
\d_{m_1\cdots m_6}^{[n_1\cdots n_6} E^{n_7n_8]n_0}  \Big)    \nn\\
\left[  E^{m_0|m_1\ldots m_8}, F_{n_0|n_1\ldots n_8}
  \right] &=& -\frac{8\cdot 8!}{9}\Big\{\big(\d^{m_0}_{n_0}\d^{m_1\ldots
  m_8}_{n_1\ldots n_8} + \d^{[m_1}_{n_0}\d^{m_2\ldots
    m_8]}_{[n_2\ldots n_8}\d^{m_0}_{n_1]}\big)K \nn\\
&&-\d^{m_1\ldots m_8}_{n_1\ldots n_8} K^{m_0}{}_{n_0} 
  - \d^{m_1\,m_2\ldots\, m_8}_{n_0[n_2\ldots n_8}K^{m_0}{}_{n_1]}^{\ } 
  - \d^{m_0[m_2\ldots m_8}_{n_1\,n_2\ldots n_8}K^{m_1]}{}_{n_0}^{\ } \nn\\
&&  - 8 \d^{m_0}_{n_0} \d^{[m_1\ldots m_7}_{[n_1\ldots n_7}
      K^{m_8]}{}_{n_8]}
   -7\d^{m_1}_{n_0}\d^{m_0m_2\ldots m_7}_{n_1n_2\ldots n_7}
     K^{m_8}{}_{n_8}\Big\}.
\eea
The normalisations with respect to the  standard bilinear form imply
\bea
\langle F_{m_1m_2m_3}| E^{n_1n_2n_3}\rangle &=& 3!\, \d^{n_1n_2n_3}_{m_1m_2m_3}
\nn\\[2mm]
\langle F_{m_1\cdots m_6}| E^{n_1 \cdots n_6}\rangle &=& 
6!\, \d^{n_1\cdots n_6}_{m_1\cdots m_6}
\nn\\[2mm]
\label{eq:ell3}
\langle F_{m_0|m_1 \cdots m_8}| E^{n_0|n_1 \cdots n_8}\rangle &=& 
\frac89 \cdot 8!\, \Big( \d^{n_0}_{m_0} \d^{n_1 \cdots n_8}_{m_1 \cdots m_8} -
      \d^{[n_1}_{m_0} \d^{n_2 \cdots n_8] n_0}_{m_1 \cdots m_8} \Big)
\eea
where the latter normalisation implies unit normalisation for real root generators 
(for which two of  the indices coincide), {\em e.g.} 
$\langle F_{1|12345678}| E^{1|12345678}\rangle = 1$.

We also recall that the finite-dimensional exceptional algebras $\lae_6, \lae_7$
and $\lae_8$ (which are of course all contained in $\lae_{10}$)
can be obtained by restricting the indices $m,n,...$ to the ranges
$\{1,...,6\}\,,\, \{1,...,7\}$  and $\{1,...,8\}$, respectively \cite{KNP}. The level expansions
for $\lae_6$ and $\lae_7$ terminate at $|\ell|=2$, while the one for $\lae_8$ extends
up to $|\ell|=3$. These truncations to finite-dimensional subalgebras of $\lae_{10}$
provide useful checks on our formulas, especially when terms must cancel 
among themselves that would otherwise be cancelled by higher level contributions
that are absent for the finite-dimensional algebras.

\subsection{Induced Actions}

We parametrise the $\E/\K$ coset element formally as
\beq\label{cV}
\cV \,=\, \cV_0 \,\cN 
\eeq
where $\cV_0 \equiv 
\exp(h^m{}_n K^n{}_m)\in {\rm GL}(10)$ corresponds to the 
standard zehnbein  $e_m{}^a$, see~\eqref{eq:elfbein}, (but as an $\E$ `matrix vielbein'), and $\cN$ to the unipotent part: 
\beq\label{cN}
 \cN = \exp\left( \frac1{3!} A_{mnp} E^{mnp} + \frac1{6!} A_{mnpqrs} E^{mnpqrs}  +
          \frac1{8!} A_{n_0|n_1\cdots n_8} E^{n_0|n_1\cdots n_8} + \cdots \right)
\eeq
Here  $m,n,p,...$ are {\em curved} (= GL(10)) indices, while $a,b,c,...$ are flat (= SO(10)) indices. 
We can use the zehnbein $e_m{}^a$ and its inverse to convert curved to flat indices on
the fields and on the $\E$ generators (recall that the fundamental form fields
$A_{mnp}, etc.$ are the ones with {\em curved} indices). When dealing with $\K$ it
is sometimes convenient to switch to flat indices on the generators, so as to be able to form
linear combinations of type $E-F$.

For the induced action we note that the above expression corresponds to a
parabolic (`almost triangular') gauge where the factor $\cV_0$ corresponds 
to the Levi subgroup GL(10), which gets completed with the unipotent 
part $\cN$ to a parabolic subgroup of $\E$. Note that with triangular $e_m{}^a$, 
the factor $\cV_0$, and with it the parabolic gauge become fully triangular.
For simplicity, and to facilitate the comparison with the standard WDW approach
we thus trade the triangular zehnbein by the metric $g_{mn}$
and its inverse.

For any wave function $\Phi = \Phi(\cV) \equiv \Phi(g_{mn}, A_{mnp},\cdots)$ we have 
the induced action
\beq
(g\circ \Phi)(\cV) \,\equiv \, \Phi \big(g^{-1} \cV k \big)
\eeq
where the compensating $\K$ transformation $k=k(g,\cV)$ is only needed 
for lower triangular $g$ to restore the parabolic gauge. To identify the differential
operators realizing the Lie algebra $\e_{10}$ we evaluate this formula for infinitesimal transformations. So we set $g_\eps = \exp (\eps X)$ to compute 
\bea\label{induced1}
(g_\eps\circ \Phi)(\cV) \,&=&\, \Phi \Big( \cV \big( {\bf 1} - \eps \cV^{-1} X \cV + \eps \d k \big) \Big)
\, + \, \cO(\eps^2)
\eea
Here we can restrict attention to the strictly upper or strictly lower triangular
transformations, as the GL(10) part is straightforward, see (\ref{gl}) below.
The differential operator realisation $\cO(X)$ of the relevant transformation is 
then obtained in the usual way as
\beq
\cO(X) \Phi(\cV) := \lim_{\eps\rightarrow 0} \, \frac1{\eps} \Big( (g_\eps \circ \Phi)(\cV) - \Phi(\cV)  \Big)
\eeq
Therefore, schematically, the procedure works as follows, where for illustrative
purposes we set the GL(10) submatrix to unity,  that is $\cV_0 = {\bf 1}$, 
and as an example consider a parabolic transformation that does not require 
compensators. For instance, with $\eps * X \equiv \frac1{3!} \eps_{mnp} E^{mnp}$,
we have
\bea
\Phi(\cN) \,&\equiv&\,  \Phi\big(A_3, A_6, A_{1,8}, \cdots \big) \; \rightarrow \nn\\[1mm]
&& \rightarrow\;
\Phi\big(A_3+\eps_3, A_6 + \eps_3 A_3, A_{1,8} + \eps_3 A_6 + \eps_3 A_3 A_3 , \cdots \big)
\eea
with appropriate (anti-)symmetrisations which we do not write out.
From this formula one can directly read off the formulas (\ref{EK}) below.
For the strictly upper triangular part the calculation of the commutators is thus 
straightforward, at least in principle, and leads to the low level results given in the following 
section. Observe that the resulting differential operators in principle
contain infinitely many terms.

The computation for the lower triangular part is more involved. Taking
as an example $X = \frac1{3!} \eps^{mnp} F_{mnp}$,  we need an infinitesimal compensator
\beq
\eps * \d k \,=\, 
 \frac1{3!} \eps^{mnp} \, \cV^{-1} \, \big(F_{mnp} - g_{mq} g_{nr} g_{ps} E^{qrs} \big)\cV
\eeq
in  (\ref{induced1}), and analogously for all $\ell\leq -2$ generators.
While upper triangular transformations push the level only up, the compensating 
transformations require levels to go up {\em and} down, though down only by 
finitely many levels at order $\cO(\eps)$, depending on the level of the $F$ 
transformation under consideration. Again this computation results in an infinite
number of terms, independently of the level.

\subsection{Differential Operators}

Applying the above procedure by including for the non-negative level differential operators the fields up to  $A_{m_0|m_1\ldots m_8}$ on $\ell=3$,
one gets the following
$\mathfrak{gl}_{10}$ generators from the induced action
\begin{align}\label{gl}
K^m{}_n &= - g_{np} \frac{\partial}{\partial g_{mp}} - \frac12 A_{np_1p_2} \frac{\partial}{\partial A_{mp_1p_2}} - \frac1{5!} A_{np_1\ldots p_5} \frac{\partial}{\partial A_{mp_1\ldots p_5}}\nn\\
&\quad - \frac1{8!} A_{n|p_1\ldots p_8} \frac{\partial}{\partial A_{m| p_1\ldots p_8}} - \frac1{7!} A_{p_0|p_1\ldots p_7 n} \frac{\partial}{\partial A_{p_0|p_1\ldots p_7 m}} + \ldots 
\end{align}
which implements the standard action of $\mathfrak{gl}_{10}$ on tensors.
Differentiation for the mixed symmetry field is normalised in the same way as the generators in~\eqref{eq:ell3}, i.e. 
\begin{align}
\frac{\partial A_{m_0| m_1\ldots m_8}}{\partial A_{n_0|n_1\ldots n_8}} = \frac89 \cdot 8!\, \Big( \d^{n_0}_{m_0} \d^{n_1 \cdots n_8}_{m_1 \cdots m_8} -
      \d^{[n_1}_{m_0} \d^{n_2 \cdots n_8] n_0}_{m_1 \cdots m_8} \Big)\,.
\end{align}
For the positive level generators up to $\ell =3$ we obtain
\begin{align}\label{EK}
E^{n_1 n_2 n_3} &\,=\, - \frac{\partial}{\partial A_{n_1 n_2 n_3}} + \frac1{12} A_{p_1p_2p_3} \frac{\partial}{\partial A_{p_1p_2p_3 n_1n_2n_3}} \nn\\
&\quad\quad + \frac1{180} A_{p_1\ldots p_6} \frac{\partial}{\partial A_{ p_1|p_2\ldots p_6 n_1n_2n_3}} -\frac1{48} A_{p_1p_2p_3} A_{p_4p_5p_6} \frac{\partial}{\partial A_{p_1|p_2\ldots p_6 n_1n_2n_3}} 
+\ldots \, \nn\\
E^{n_1\ldots n_6} &\,=\,  - \frac{\partial}{\partial A_{n_1\ldots n_6}}  + \frac1{12} A_{p_1p_2p_3} \frac{\partial}{\partial A_{p_1|p_2p_3 n_1\ldots n_6}} + \ldots  \, \\
E^{n_0| n_1\ldots n_8} &\,=\, -\frac{\partial}{\partial A_{n_0| n_1\ldots n_8}} +\ldots \,.\nn
\end{align}
For the negative level generators we only give the generators at levels $\ell =-1,-2$ and only include 
the contributions from the coordinates up to $A_{n_1\ldots n_6}$ since the expressions become 
very unwieldy for $\ell < -2$. Since we have not performed the dualisation of gravity in 
$D=11$ supergravity, the expressions are sufficient for the results of this paper. We get {\allowdisplaybreaks
\begin{align}
F_{n_1n_2n_3} &= 3 g_{p[n_1} A_{n_2n_3]q} \frac{\partial}{\partial g_{pq}} -\frac13 A_{n_1n_2n_3} g_{pq} \frac{\partial}{\partial g_{pq}}\\*
&\quad - g_{n_1p_1} g_{n_2p_2} g_{n_3p_3} \frac{\partial}{\partial A_{p_1p_2p_3}}  -\frac16 A_{n_1n_2n_3p_1p_2p_3} \frac{\partial}{\partial A_{p_1p_2p_3}}\nn\\*
&\quad -\frac1{12} A_{n_1n_2n_3} A_{p_1p_2p_3} \frac{\partial}{\partial A_{p_1p_2p_3}} +\frac34 A_{p_1[n_1n_2} A_{n_3]p_2p_3} \frac{\partial}{\partial A_{p_1p_2p_3}} \nn\\*
&\quad + \frac1{12}g_{n_1p_1} g_{n_2p_2} g_{n_3p_3} A_{p_4p_5p_6} \frac{\partial}{\partial A_{p_1\ldots p_6}}-\frac1{720} A_{n_1n_2n_3} A_{p_1\ldots p_6} \frac{\partial}{\partial A_{p_1\ldots p_6}}\nn\\*
&\quad +\frac1{80} A_{p_1\ldots p_5[n_1} A_{n_2n_3]p_6}  \frac{\partial}{\partial A_{p_1\ldots p_6}} +\frac{1}{48} A_{p_1p_2p_3} A_{p_4p_5[n_1} A_{n_2n_3]p_6}  \frac{\partial}{\partial A_{p_1\ldots p_6}}\nn\\*
& \quad + \, \cdots \nn\\
F_{n_1\ldots n_6} &= - 6 g_{p[n_1} A_{n_2\ldots n_6]q} \frac{\partial}{\partial g_{pq}} -  \frac23 A_{n_1\ldots n_6} g_{pq} \frac{\partial}{\partial g_{pq}}- 30 g_{p[n_1} A_{n_2n_3n_4} A_{n_5n_6]q} \frac{\partial}{\partial g_{pq}}\nn\\*
&\quad -\frac53 A_{s_1s_2s_3[n_1n_2n_3} A_{n_4n_5n_6]} \frac{\partial}{\partial A_{s_1s_2s_3}} -\frac32 A_{s_1s_2[n_1} A_{n_2\ldots n_6]s_3} \frac{\partial}{\partial A_{s_1s_2s_3}} \nn\\*
&\quad -\frac16 A_{s_1s_2s_3} A_{n_1\ldots n_6} \frac{\partial}{\partial A_{s_1s_2s_3}} -5 A_{s_1s_2[n_1} A_{n_2n_3n_4} A_{n_5n_6]s_3} \frac{\partial}{\partial A_{s_1s_2s_3}}\nn\\*
&\quad -20 A_{n_1n_2n_3} g_{n_4s_1}g_{n_5s_2}g_{n_6s_3}\frac{\partial}{\partial A_{s_1s_2s_3}}\nn\\*
&\quad -\frac{3}{5!} A_{n_1s_1\ldots s_5} A_{s_6n_2\ldots n_6} \frac{\partial}{\partial A_{s_1\ldots s_6}} - \frac1{3\cdot 5!} A_{n_1\ldots n_6} A_{s_1\ldots s_6}\frac{\partial}{\partial A_{s_1\ldots s_6}}  \nn\\*
&\quad + \frac1{12} A_{n_1s_1\ldots s_5} A_{s_6n_2n_3} A_{n_4n_5n_6} \frac{\partial}{\partial A_{s_1\ldots s_6}} -\frac{5}{108} A_{s_1s_2s_3} A_{s_4s_5s_6n_1n_2n_3} A_{n_4n_5n_6} \frac{\partial}{\partial A_{s_1\ldots s_6}} \nn\\*
&\quad +\frac{5}{24} A_{s_1s_2s_3} A_{s_4n_1\ldots n_5} A_{n_6s_5s_6} \frac{\partial}{\partial A_{s_1\ldots s_6}} -\frac{5}{24} A_{s_1s_2s_3} A_{s_4s_5n_1} A_{n_2n_3n_4} A_{n_5n_6s_6} \frac{\partial}{\partial A_{s_1\ldots s_6}} \nn\\*
&\quad - g_{n_1s_1} \cdots g_{n_6s_6} \frac{\partial}{\partial A_{s_1\ldots s_6}} - \frac53 A_{s_1s_2s_3} A_{n_1n_2n_3} g_{n_4s_4} g_{n_5s_5} g_{n_6s_6} \frac{\partial}{\partial A_{s_1\ldots s_6}} 
\nn\\*
& \quad + \, \cdots
\end{align}
Even when not writing out antisymmetrisations explicitly, all terms on the right-hand sides are understood
to be antisymmetrised properly in the $n_i$ indices. }

Up to $\lae_7$ for which the indices assume only the values $m,n,...\in \{1,...,7\}$, the level decomposition stops at this level and one can check using Schouten identities that all commutators close correctly. For higher rank $\lae_n$ this requires contributions from yet higher levels that we have not worked out.

We also note that the property that $E$ and $F$ are each other's Hermitian conjugates 
is not manifest in this functional realisation. In principle, this requires an appropriate
measure on the function space that the differential operators are acting on. To the best 
of our knowledge, such a measure is not known for the symmetric space $\E/\K$,
but does exist for the finite-dimensional truncations.
In section~\ref{sec:qBKL} below, we shall investigate the measure for finite-dimensional subspaces.

\subsection{\texorpdfstring{WDW Hamiltonian and the $\E$ Casimir operator}{WDW Hamiltonian and the E10 Casimir operator}}
\label{sec:comp}

As is well known~\cite{Kac}, one can define a normal ordered Casimir operator for $\E$ 
when acting on integrable modules with a highest weight element. The $\E$ Casimir with parabolic normal 
ordering adapted to the $GL(10)$ decomposition is 
\begin{align}
\label{WDW2}
\Omega = \frac12 K^m{}_n K^n{}_m -\frac1{18} KK +\frac{23}{6} K + \frac1{3!} F_{mnp} E^{mnp} 
          + \frac1{6!} F_{m_1\ldots m_6} E^{m_1\ldots m_6} + \cdots  
\end{align}
The term linear in $K$ is fixed by requiring the Casimir to commute with $\E$. 
To determine its coefficient it is enough to check that $\Omega$ commutes 
with $E^{mnp}$ and $F_{mnp}$, as all the higher level generators are given 
by multi-commutators of the ones at $\ell = \pm 1$; the remaining terms are 
uniquely given by the general expression (\ref{Casimir}).
The above expression differs from the standard one \cite{Kac} only in its level 
zero contribution, where instead there appears a contribution depending on the 
$\lae_{10}$ Weyl vector $\varpi$. This 
difference is the result of a partial reordering of the $\ell =0$ generators, 
since the $\mathfrak{gl}_{10}$ generators 
on level $\ell=0$ are not normal ordered, unlike the $\ell\neq 0$ terms. Taking this
difference into account we have complete agreement with the standard formula,
as we will demonstrate in appendix~\ref{app:Cas}.

Now let us work out the Casimir up to $\ell =2$ with our explicit expressions for
the $\E$ generators. As it turns out there are numerous cancellations,
and after some algebra we are left with\footnote{The $\text{E}_n$ Casimir 
for $n\leq 8$ was already worked out in~\cite{Obers:1999um} in the same truncation.}
\begin{align}
\label{eq:Om10}
\Omega &= \frac12 \left[ g_{mp} \frac{\p}{\p g_{np}} g_{nq} \frac{\p}{\p g_{mq}}
                      - \frac19 \left( g_{mn} \frac{\p}{\p g_{mn}} \right)^2 \right]
                      - \frac{23}6  g_{mn} \frac{\p}{\p g_{mn}}   \\[2mm]
 &\quad + \frac1{3!} \, g_{mq} g_{nr} g_{ps} \frac{\p}{\p A_{mnp}}\frac{\p}{\p A_{qrs}} \,+\,
 \frac1{6!} \, g_{m_1n_1}\cdots g_{m_6n_6} \frac{\p}{\p A_{m_1\cdots m_6}}
  \frac{\p}{\p A_{n_1\cdots n_6}} \nn\\[2mm]
&\quad - \frac{1}{36}\,
g_{mq_1} g_{nq_2} g_{pq_3} A_{q_4q_5 q_6}
         \frac{\p}{\p A_{q_1\cdots q_6}} \frac{\p}{\p A_{mnp}} \nn\\[2mm]
&\quad +  \frac1{3!} \cdot \frac1{144}  \, g_{k_1 p_1} g_{k_2 p_2}g_{k_3 p_3} A_{m_1m_2m_3}
A_{n_1n_2n_3} \frac{\pa}{\pa A_{m_1m_2m_3k_1k_2k_3}} \frac{\pa}{\pa A_{n_1n_2n_3p _1p_2 p_3}}  \,       
 \,+\,\ldots    \nn
 \end{align}
Remarkably many cross terms cancel, in particular the ones 
$\propto \d/\d g \cdot \d/\d A$, and it is an interesting question whether such 
cancellations still persist beyond level $\ell =2$. 
We also notice that the terms involving the three- and six-form
variables can be written more simply and more suggestively as 
\beq\label{OmFF}
\Omega \,\Big|_{|\ell| =1,2} \,=\,  \frac{e^2}{4!} \, g^{m_1n_1} \cdots g^{m_4n_4} \cF_{m_1\cdots m_4}\cF_{n_1\cdots n_4}
\,+\, \frac{e^2}{7!} \, g^{m_1n_1} \cdots g^{m_7n_7} \cF_{m_1\cdots m_7}\cF_{n_1\cdots n_7}
\eeq
where 
\begin{align}
\cF_{m_1\ldots m_4} &= -\frac{1}{6!} \ve_{m_1\ldots m_4n_1\ldots n_6}  \frac{\partial}{\partial A_{n_1\ldots n_6}} \,,\nn\\
\cF_{m_1\ldots m_7} &= -\frac{1}{3!} \ve_{m_1\ldots m_7n_1\ldots n_3} \left( \frac{\partial}{\partial A_{n_1\ldots n_3}} + \frac1{12} A_{s_1s_2s_3} \frac{\partial }{\partial A_{s_1s_2s_3n_1n_2n_3}}\right)\,.
\end{align}

Comparison of (\ref{OmFF}) with (\ref{eq:QH0}) now immediately shows that, up to and 
including level $\ell=2$, this structure  coincides with the bosonic Hamiltonian 
of $D=11$ supergravity. That is, {\em at a given spatial point}, we have the equality
\beq\label{H=Om}
e \hat\cH_0 \,=\, -  \frac{2 \hbar^2 \GN}{\ell_{\rm P}^{\,20}}
\,\Omega \, \Big|_{|\ell|\leq 2}
\eeq
provided we convert the functional differential operators
into ordinary partial derivatives according to the rule (\ref{FDer}).
We also note that formula (\ref{H=Om}) is consistent with the fact that $\hat{\cH}_0$ 
has dimension $M\cdot L^{-10}$ (= energy density) while $\Omega$ is dimensionless. We recall that $\ell_{\rm P}= (\hbar \GN)^{1/9}$ and so only two independent fundamental constants appear in~\eqref{H=Om}.

We can no longer expect complete matching between the $\E$ Casimir and the 
$D=11$ Hamiltonian beyond $\ell=2$ without a proper dualisation of gravity. 
This is already clear from the
absence of the trace of the spin connection $\omega_{bb a}$ on the 
$\E$ side, and from the fact 
that the positivity of the $\E$ Casimir away from the Cartan subalgebra is in 
conflict with the fact that the spatial curvature contribution $\propto R^{(10)}$
in the WDW Hamiltonian (\ref{eq:QH0}) can have either sign. There are further
mismatches at $\ell = 3$ which were already exposed in \cite{Cocoyoc,Damour:2006xu}.
At yet higher levels, the known correspondence (`dictionary') breaks down altogether.

Nevertheless, disregarding the remaining discrepancies, we note that (\ref{eq:Om10})
has a fixed ordering of the differential operators which is uniquely prescribed 
by the form of the $\E$ Casimir, with all  differential operators to the right, except for 
standard WDW term in the first line (in cancelling contributions proportional to 
$A_3 \,\pa/\pa A_3$ and $A_6 \, \pa/\pa A_6$ the term $\frac{23}6 K$ again
plays a crucial role). We also note that by the very definition of the Casimir
operator we have an infinite number of $\E$ `charges' that commute with the
Casimir, namely all operators corresponding to the $\E$ generators. Whether
these admit a space-time interpretation as `observables' remains to be explored.

With the identification (\ref{H=Om})  the WDW operator acquires a `dimensionless' 
form since all terms in~\eqref{eq:Fop} are homogeneous in dimensionful constants after dropping the terms involving spatial gradients.
In this form the semi-classical limit $\hbar\ra 0$ evidently cannot be
meaningfully discussed, as $\hbar$ appears only as an overall factor.  
The issue of the semi-classical limit is thus intimately connected to the
question of space emergence; the requisite dimensionful parameters
only appear after inclusion of the spatial derivative terms.

Let us also comment on the remaining constraints. As emphasised in section~\ref{sec:11D}, these involve spatial derivatives, see for instance~\eqref{eq:diff} and~\eqref{eq:G2G5}. 
Applying the same substitutions of magnetic field strengths~\eqref{C0} (or more properly their second-class version) to the terms not involving explicit spatial derivatives, the first Gauss constraint~\eqref{eq:G2G5} becomes for example 
\begin{align}
\cG^{mn} &\,=\, \partial_p \Pi^{mnp} + \frac1{864}  \ve^{mn q_1\ldots q_8}F_{q_1\ldots q_4} F_{q_5\ldots q_8} \nn\\
&\to  \;\partial_p \frac{\d}{\d A_{mnp}} 
- \frac1{36 \cdot 6!} \ve_{q_1\ldots q_{10}} \frac{\d}{\d A_{mnq_1\ldots q_4}} \frac{\d}{\d A_{q_5\ldots q_{10}}} +\ldots
\end{align}
where we have applied our quantisation from section~\ref{sec:quant} and so converted the constraint into a  functional differential operator in field space up to the explicit spatial derivative. The ellipses include terms involving spatial gradients of the fields and non-local terms that are due to solving the second-class constraints. 
We expect that similar manipulations can be applied to the diffeomorphism constraint after dualising gravity at the linearised level. These differential operators have to be applied to the WDW wave functional $\Psi$. In the comparison to $\E$ we drop the explicit spatial 
derivatives and non-local terms and arrive at an ultra-local expression that can also be interpreted as 
a constraint on the $\E$ wave function $\Phi$ if one transitions according 
to~\eqref{Psi}.
 This represents the quantum version of the classical constraints studied in~\cite{DKN,Damour:2009ww} that can be imposed consistently on the classical $\E$ model. We note that, when determining the $\lae_{10}$ weight of the components of the constraints, there seems to be a relation to the indecomposable extension of $\lae_{10}$ studied in~\cite{BKS}.

\section{Comparison with quantum BKL analysis}
\label{sec:qBKL}

In this section, we consider the solutions of the $\E$ WDW equation $\Omega\Phi =0$ and their relation to previous work on the quantisation of the BKL/cosmological billiards approximation to $D=11$ supergravity in the vicinity of a space-like singularity~\cite{Belinsky:1970ew,Damour:2000hv,Damour:2002et,Henneaux:2007ej}. The quantisation of the cosmological billiards picture was found to lead to a normalisable wave function of the Universe that tends to zero when approaching the singularity~\cite{Kleinschmidt:2009cv}, thus realising DeWitt's original idea of the quantum mechanical resolution of classical singularities~\cite{DeWitt}, see also~\cite{Misner:1969hg,Graham:1990jd,Damour:2013eua,Perry:2021mch}. We shall review this result that uses only the Cartan subalgebra of $\mathfrak{e}_{10}$ together with the input of the walls locations from cosmological billiard. We then generalise the analysis to include also root generators but we restrict mainly to the case of a single root generator for simplicity. This does not alter the physical conclusions and also connects to the idea of discrete symmetries in string theory.

In group theoretical terms, the BKL approximation corresponds to the restriction 
of $\lae_{10}$ to its Cartan subalgebra $\lah$. Here we shall generalise this setting
to larger finite-dimensional subalgebras of $\lae_{10}$
\beq\label{BKL}
\lah_\perp \oplus \lag_{(r)} \,\subset \, \lae_{10} \;\;,\quad
\langle \lah_\perp | \lag_{(r)} \rangle \,=\, 0 \;,\quad
d \equiv {\rm dim} \,\lah_\perp = 10 - r
\eeq
corresponding to a compactification from $D=11$ down to $D=d+1$ space-time dimensions.
Here, the subalgebra $\lag_{(r)}$ is of rank $r$, and the dimension $d$ of the restricted 
Cartan subalgebra $\lah_\perp\subset \lah$ coincides with the dimension 
of the singular spatial hypersurface. The relevant modular group is then 
$\cW_\perp  \times {\rm G}_{(r)}(\mathbb{Z})$, where $\cW_\perp$ 
is the even subgroup of the Weyl group associated to the billiard defined by 
the remaining walls for $\lah_\perp$, and  ${\rm G}_{(r)}(\mathbb{Z})$
the appropriate discrete U-duality group for the matter sector. Below we will in particular 
consider the cases $\lag_{(1)} \equiv \lag_\a = \mathfrak{sl}_2$ for arbitrary real and 
timelike imaginary roots $\a$, as well as higher rank examples such as
$\lag_{(7)} = \lae_7$ and $\lag_{(8)} = \lae_8$.

\subsection{Review of quantum cosmological billiards}

In~\cite{Kleinschmidt:2009cv,Kleinschmidt:2009hv}, the quantisation of the $\E$ cosmological billiard was studied. The analysis  is based on the mini-superspace approximation where only the diagonal components of the spatial metric are retained and their free motion is constrained by hard walls that are the only remnant of the other components and the matter fields. The ten diagonal components of the metric are associated with the Cartan subalgebra generators $K^m{}_m$ for $m=1,\ldots,10$ (no sum). As the DeWitt metric~\eqref{eq:DWM} reduces to a Lorentzian metric $\eta_{mn}$ of signature $(1,9)$ on diagonal metrics, a convenient set of coordinates for the diagonal components 
 $g_{mm} = \exp(-2\beta^m)$ is given by \cite{Damour:2002et}
\begin{align}
\label{eq:BKLvars}
\beta^m = \rho \gamma^m \;, \qquad \mbox{with} \quad
\rho>0 \quad \text{and}\quad \gamma^n \eta_{mn} \gamma^n = -1
\end{align}
with the logarithmic scale factors $\beta^m$, and
coordinates $\gamma^m$ on the unit hyperboloid and $\rho$ representing the effective time parameter for the approach to the singularity which is at $\rho\to \infty$ in these coordinates. Together, $\rho$ and $\gamma^m$ parametrise the interior of the forward light-cone in the space of diagonal spatial metrics. 
The hard `billiard' walls constrain the motion in the forward light-cone to a fundamental chamber of the $\E$ Weyl group~\cite{Damour:2000hv}. Quantum-mechanically, one has to solve the wave equation on this Lorentzian space with Dirichlet boundary conditions corresponding to the hard walls.\footnote{For a discussion of other boundary conditions and related ideas see~\cite{Misner:1969hg,Graham:1990jd,Forte:2008jr}.} 

The wave operator is invariant under the $\E$ Weyl group and given by
\begin{align}
 \frac12 \sum_{m=1}^{10} \left( K^m{}_m \right)^2 - \frac1{18} \left( \sum_{m=1}^{10}K^m{}_m \right)^2  = - \rho^{-9} \partial_\rho ( \rho^9 \partial_\rho)  + \rho^{-2} \Delta_{\text{LB}}\,,
\end{align}
where we now suspend the summation convention by writing out sums explicitly.
Here $\Delta_{\text{LB}}$ is the Laplace--Beltrami operator on the unit hyperboloid of dimension nine.
These terms are recognised as the restriction of the $\E$ Casimir~\eqref{WDW2} to the Cartan generators, \textit{except} for the normal-ordering term $\frac{23}{6} K$. The latter breaks $\E$ Weyl symmetry on its own. However, together with all root generators in~\eqref{WDW2} (full, continuous) $\E$ symmetry is restored. 
The restriction of the full wave equation $\Omega \Phi=0$ to dependence only on diagonal metric components therefore differs from the wave equation coming from the cosmological billiard (with boundary conditions given by the hard walls) by the normal-ordering term. One may wonder whether the addition of this term will modify the conclusion of~\cite{Kleinschmidt:2009cv} regarding the vanishing of the wave function at the singularity.

In order to investigate this we recall from~\cite{Kleinschmidt:2009cv} that the spectrum of the Laplace--Beltrami operator on the unit hyperboloid in $d=10$ dimensions (with Dirichlet boundary conditions) is bounded by
\begin{align}
\label{eq:LBspec}
-\Delta_{\text{LB}} \geq 16\,,
\end{align}
which, together with a separation ansatz 
$\Phi(\rho,\gamma^m) = \cR(\rho) \Phi_0(\gamma^m)$
for the wave function leads to the result that
\begin{align}
\label{eq:qBKL}
\Phi(\rho,\gamma^m)  \sim \rho^{-4} e^{i \mu \log \rho} \Phi_0 (\gamma^m)\,,
\end{align}
where $(- \Delta_{\text{LB}} -16)\Phi_0 = \mu^2 \Phi_0$ and the
reality of $\mu$ is guaranteed by~\eqref{eq:LBspec}. 
Therefore the full wave function vanishes (and oscillates) for $\rho\to \infty$. 

For later reference we recall that for singular spatial hypersurfaces of 
dimension $d$, the relevant operator is 
$ -\rho^{1-d} \pa_\rho (\rho^{d-1}\pa_\rho) + \rho^{-2} \Delta_{\text{LB}}$.
Then the bound in (\ref{eq:LBspec}) becomes 
 \beq\label{eq:Bspec0}
  -\Delta_{\text{LB}} \geq \frac14 (d-2)^2
\eeq
and the wave function decays as $\rho^{-(d-2)/2}$~\cite{Kleinschmidt:2009cv}.
The conserved invariant measure for the quantum cosmological billiards is 
just given by the standard Klein--Gordon inner product 
\beq\label{KG}
(f | g ) \,=\, i \int d \Sigma^a f^* \stackrel{\leftrightarrow}{\pa_a} g
\eeq
where the integral is over the unit hyperboloid inside the forward light-cone in 
$\beta$-space.
We stress that these statements are true for $D=11$ supergravity, but similar results hold for other gravitational theories such as pure $D=4$ gravity without a cosmological constant.

\subsection{Extension with a single root}
 
Since considering the full $\E$ system is too complicated (and possibly hard to define properly), we consider the case when only a single positive root generator $E_{\alpha}$ is active along with its associated negative root generator in addition to the Cartan subalgebra. This means that the algebra we are considering is
\begin{align}
\label{eq:hplus1}
\mathfrak{h}_\perp \oplus \mathfrak{g}_\alpha \subset \lae_{10}
\,,\quad \quad  \lag_{(1)} \equiv
\mathfrak{g}_\alpha = \{ E_\alpha , F_\alpha , H_\alpha \equiv \alpha^i H_i  \}\,, \quad\quad
\langle \mathfrak{h}_\perp | \mathfrak{g}_\alpha \rangle = 0\,.
\end{align}
whence the $\lae_{10}$ symmetry is broken to 
$\mathfrak{h}_\perp \oplus \mathfrak{g}_\alpha$. Physically,
this truncation corresponds to a situation with one compactified dimension, where
the dimension of the singular spatial hypersurface is reduced by one. 

The three-dimensional algebra $\mathfrak{g}_\alpha$ is isomorphic (over $\mathbb{R}$) to $\mathfrak{sl}_2(\mathbb{R})$ if $\alpha^2\neq 0$\footnote{The induced bilinear form has non-standard signature 
for time-like $\alpha$.} and isomorphic to a Heisenberg algebra if $\alpha$ is a null root. 
The direct sum in the Lie algebra~\eqref{eq:hplus1} is one of Lie algebras. We use the normalisation
\begin{align}
\label{eq:norm}
\langle E_\alpha | F_\alpha \rangle  =1 \,,\quad 
\langle H_\alpha| H_\alpha \rangle = \alpha^2\,.
\end{align}
This bilinear form is invariant under the commutation relations
\begin{align}
[ H_\alpha, E_\alpha ] = \alpha^2 E_\alpha\,,\quad
[ H_\alpha, F_\alpha ] = - \alpha^2 F_\alpha\,,\quad
[ E_\alpha, F_\alpha ] = H_\alpha\,.
\end{align}
Classical cosmological solutions to the $\E$ sigma model in such a set-up have 
been studied in~\cite{Kleinschmidt:2005gz}, but we are here interested in the quantisation.

The Casimir operator for the algebra~\eqref{eq:hplus1} is likewise
a truncation of the full $\E$ Casimir operator to a finite-dimensional
differential operator, and decomposes as
\begin{align}
\Omega_1 = \Omega_\perp + \Omega_\alpha\,,
\end{align}
where $\Omega_\perp$ is the part along $\mathfrak{h}_\perp$ and $\Omega_\alpha$ along $\mathfrak{g}_\alpha$. 
From now on we only consider the case $\alpha^2\neq 0$ such that the bilinear form~\eqref{eq:norm} is non-degenerate and the Casimir reads
\begin{align}
\Omega_\alpha =  \frac12 E_\alpha F_\alpha + \frac12 F_\alpha E_\alpha + \frac1{2\alpha^2} H_\alpha H_\alpha = F_\alpha E_\alpha +\frac12 H_\alpha + \frac1{2\alpha^2} H_\alpha H_\alpha \,.
\end{align}

Choosing coordinates $\chi$ and $\phi$ on an Iwasawa patch of the symmetric space associated with $\mathfrak{g}_\alpha$ with representative $\cV_\alpha = e^{\chi E_\alpha} e^{\phi H_\alpha}$ we have the differential operators
\begin{align}
\label{eq:dops}
E_\alpha &= - \partial_\chi \,,\quad H_\alpha = -\partial_\phi -\alpha^2 \chi \partial_\chi\,,\nn\\
F_\alpha &= \chi \partial_\phi + \left( \frac12 \alpha^2 \chi^2 -e^{2\alpha^2 \phi} \right) \partial_\chi
\end{align}
and therefore
\begin{align}
\label{eq:Omalpha}
\Omega_\alpha &= \frac12 \alpha^{-2} \partial_\phi^2 -\frac12 \partial_\phi+ e^{2\alpha^2 \phi} \partial_\chi^2
= \frac12 \alpha^2 y^2 \partial_y^2 + y^2 \partial_\chi^2\,,
\end{align}
where we have defined $y= e^{\alpha^2 \phi}$ in the last step to make the expression 
coincide with the usual $SL(2,\mathbb{R})$ Laplace operator on the upper half-plane 
for real roots ($\alpha^2=2$).

In order to solve the WDW equation 
\begin{align}
\label{eq:WDW1}
\Omega_1 \Phi_1 ( \beta_\perp, \phi,\chi)  =0 
\end{align}
we use separation of variables, with
$\Phi(\beta_\perp,\phi,\chi) = \Phi_\perp(\rho,\gamma) \cF(\phi,\chi)$
as well as $\Phi_\perp (\rho,\gamma) =  \cR(\rho) \Phi_0(\gamma)$. 
If $\Omega_\alpha \cF(\phi,\chi) = - \mathcal{E} \cF(\phi,\chi)$ we are left with
\begin{align}
\label{eq:WDWperp}
\Omega_\perp \Phi_\perp(\beta_\perp) = \mathcal{E} \Phi_\perp (\beta_\perp)\,.
\end{align}
We analyse the case of real roots and time-like imaginary roots separately.

We note that for all $\alpha^2\neq 0$ the integration measure on the homogeneous space associated with $\mathfrak{g}_\alpha$ the integration measure is given by
\begin{align}\label{Measure}
(f | g ) = \sqrt{\frac{2}{|\alpha^2|}}\int \frac{d\chi dy}{y^2} f(\chi,y) g(\chi,y)
\end{align}
and the operators~\eqref{eq:dops} all satisfy $X^\dagger = - X$ with respect to this integration measure. The constant overall normalisation factor is conventional 
and could be dropped. This measure supplements the billiard measure (\ref{KG})
to provide a measure for the full wave function $\Phi_1(\beta_\perp,\phi,\chi)$.
What is important is that the patch where the coordinates $\chi\in\mathbb{R}$ and $y>0$ are defined is of infinite volume in the measure (\ref{Measure}).
This can be remedied by considering a quotient of the homogeneous space by a discrete subgroup such that quotient has finite volume. A standard example for this is the modular group $SL(2,\mathbb{Z})$ acting on the upper half-plane which makes contact of the present set-up to the theory of automorphic forms. As mentioned in the introduction, such discrete symmetries arise naturally in an M-theory context, and here we see a different need for them in quantum gravity.

\subsection{The case of a real root}

If $\alpha$ is a real root, the associated symmetric space is $SL(2,\mathbb{R})/SO(2)$, the two-dimensional hyperbolic plane. The orthogonal space $\mathfrak{h}_\perp$ is a Lorentzian space of dimension $9$ and we can choose coordinates $\beta_\perp = \rho \gamma_\perp$ similar to~\eqref{eq:BKLvars}. Separating the equation~\eqref{eq:WDWperp}, we find a total elementary solution to~\eqref{eq:WDW1} of the form 
\begin{align}
\label{eq:Phi1}
\Phi_1( \rho, \gamma_\perp, \phi,\chi) = \rho^{-\frac72} \cR(\rho) \Phi_0(\gamma_\perp) \cF(\phi,\chi)
\end{align}
with $\Delta_{\rm LB}  \Phi_0 (\gamma_\perp) = - E \Phi_0(\gamma_\perp)$ on the eight-dimensional 
unit hyperboloid inside $\mathfrak{h}_\perp$ and $\cR(\rho)$ a bounded function of $\rho$ that 
is given by the solution to an (ordinary) Bessel differential equation.
More precisely, $\cR(\rho)$ satisfies the equation
\begin{align}
\rho^2 \partial_\rho^2 \cR + \rho \partial_\rho \cR + \left(\mathcal{E} \rho^2
 + E-\frac{49}{4}\right) \cR = 0\,,
\end{align}
where the contribution $\frac{49}4$ corresponds to $d=9$ in (\ref{eq:Bspec0}).
After rescaling $\mathcal{E}\rho^2 \to  \rho^2$ this becomes the standard Bessel 
equation, but with {\em imaginary} index. An explicit solution is provided by 
the formula \cite{Bessel}
\beq
\cR (\rho) = 
J_\nu(\mathcal{E}^{-1/2}\rho) \,=\, \frac{2(\frac12 \mathcal{E}^{-1/2}\rho)^{\nu}}{\sqrt{\pi} \Gamma(\nu + \frac12)} \int_0^1 (1 -t^2)^{\nu -\frac12}
     \cos(\mathcal{E}^{-1/2} \rho t) dt  
\eeq
valid for Re$(\nu) > -\frac12$. The index
\beq
\nu = i \sqrt{E- \frac{49}4}\;. 
\eeq
is purely imaginary because $E\geq \frac{49}{4}$, which is the appropriate bound 
for the spectrum of the Laplacian $-\Delta_{\rm LB}$ for $d=9$ by (\ref{eq:Bspec0}),
under the assumption that there is a restricted cosmological billiard with Dirichlet boundary conditions. Moreover, we have that $\mathcal{E}\geq \frac14$ under the assumption of a duality symmetry acting on $SL(2,\mathbb{R})/SO(2)$ with Dirichlet boundary conditions for normalisable solutions $\cF(\phi,\chi)$. 
This bound makes the rescaling of the variable real. In the range 
\begin{align}
\label{eq:range}
E \geq \frac{49}{4} \quad\text{and}\quad \mathcal{E} \geq \frac14
\end{align}
we then have that $\Phi_1$ is normalisable. In addition, the Bessel function $\cR(\rho)$ is complex, 
oscillating and decays as $\rho^{-1/2}$ for $\rho\to \infty$ in the same range of parameters. 
Therefore, the full wave function $\Phi_1$ retains the property  that it vanishes 
in an oscillating manner for $\rho \to \infty$ when one simple real root is turned on.

The variable $\rho$ used in~\eqref{eq:Phi1} is the one associated only to the space $\mathfrak{h}_\perp$ rather than to all of $\mathfrak{h}$ as in~\eqref{eq:qBKL}. The reason for this change is that turning on the root generator $E_\alpha$ and the associated variable $\chi$ removes one hard billiard wall and the BKL geometry 
is therefore changed, with a singular spatial hypersurface of dimension $d=9$.

Let us also recall that there are {\em infinitely} many real roots for $\E$ (like for other
hyperbolic Kac-Moody algebras), whose associated root generators can be used 
to build what is often referred to as the `minimal group', corresponding to a special
prescription for exponentiating the Lie algebra $\lae_{10}$ \cite{KacPeterson}.
While we have no general statement about the behavior of the full $\E$ wave function,
this observation already takes us some way towards establishing the generic
vanishing of the wave function at the space-like singularity.

\subsection{Automorphic aspects with several real roots}

For deriving the range~\eqref{eq:range}, we used that there are discrete symmetries acting on the variables. On the space $\mathfrak{h}_\perp$ the discrete symmetry was the remnant of the $\E$ Weyl symmetry. This is an infinite order symmetry since the stabiliser of any real root inside the infinite $\E$ Weyl group is of finite order. 

For the $SL(2,\mathbb{R})/SO(2)$ symmetric space associated with the real root $\alpha$ we assumed a discrete symmetry such as $SL(2,\mathbb{Z})$ with a fundamental domain of finite volume, together with Dirichlet conditions.\footnote{Under these conditions, the bound $\mathcal{E}\geq \tfrac14$ can actually be strengthened~\cite{Terras}, but we do not require this here. In fact, all that we require is that $\mathcal{E}$ is non-negative.} The mathematical reason is that under these assumptions the derivation of the bound on $\mathcal{E}$ is straight-forward~\cite{Iwaniec} while without this assumption the bound is only almost always satisfied~\cite{Lax:1982}. More importantly, the physical reason for this assumption is that in string theory and supergravity, maximal supersymmetry together with Dirac charge 
quantisation implies the existence of such discrete U-duality groups~\cite{Hull:1994ys} 
that are associated with space-time, at least for $D = d+1 \geq 4$.

\medskip

The arguments above can be extended to the case when $\mathfrak{g}_\alpha$ is replaced by any finite-dimensional subalgebra $\mathfrak{h}_\perp \oplus \mathfrak{g}\subset \mathfrak{e}_{10}$, e.g. $\mathfrak{g}= \mathfrak{e}_7$ in which case $\mathfrak{h}_\perp$ would be of dimension three. For finite-dimensional $\lag \equiv \mathfrak{g}_{(r)}$ 
semi-simple of rank $r$ such that $\dim\mathfrak{h}_\perp=10\!-\!r$, the same separation ansatz 
\begin{align}
\Omega_r = \Omega_\perp + \Omega_{\mathfrak{g}}
\end{align}
applies, and we can first solve $\Omega_{\mathfrak{g}} \cF = - \mathcal{E} \cF$. The appropriate generalisation of the measure~\eqref{Measure} exists on the finite-dimensional symmetric space associated with $\mathfrak{g}$ by standard results, i.e., computing the invariant metric and taking its determinant. For the unipotent part associated with the positive roots this is given by the usual Haar measure.

For instance, for $\lag = \lae_7$
this corresponds to 
a situation with a singular spatial hypersurface of dimension three, and seven compactified dimensions.
For doing this, we assume again the existence of discrete U-duality and require the square integrable functions 
on a locally symmetric space. Although the precise general bound on the Laplace spectrum is not known 
for E$_7(\mathbb{Z})$ to the best of our knowledge, one can still show that $\mathcal{E}\geq 0$. 
One instance of a such a function $\cF$ for $E_7$ or $E_8$ is provided 
by the automorphic realisation of the minimal unitary 
representation~\cite{Kazhdan:2001nx,Gunaydin:2001bt,Green:2011vz,Fleig:2015vky}.

For $\mathcal{E}>0$ we are then led again to solutions of the type~\eqref{eq:Phi1} where the difference is that $\rho^{-7/2}$ is replaced by $\rho^{-(8-r)/2}$ and the solution to the Bessel equation provides an additional falloff, still ensuring that the wave function of the Universe vanishes when approaching the singularity as a stable property of the solution. Here, we assume implicitly that $r\leq 8$ to have a meaningful geometric picture of the singularity when using the dictionary.

\subsection{The case of an imaginary root}
\label{sec:imroot}

For (time-like) imaginary roots $\a$ (which obey $\a^2 < 0$) new subtleties arise. A first
difficulty is that the `dictionary' of \cite{Damour:2002cu,Cocoyoc} does
not work for imaginary roots, hence there does not exist an obvious 
geometric interpretation for this situation, unlike for cosmological billiards.
While for real roots the geometry of the coset is that of $SL(2,\mathbb{R})/SO(2)$, 
the symmetric space associated with an imaginary root is $SL(2,\mathbb{R})/SO(1,1)$,
which is now of Lorentzian signature. For $\a^2 = -2$ (or any time-like root 
after a suitable rescaling of the coordinates), the metric reads
\begin{align}\label{AdS}
\dd s^2 = y^{-2} \left( -\dd y^2 + \dd\chi^2\right)
\end{align}
and represents a Poincar\'e patch of Lorentzian AdS$_2$ space that 
was computed using one Iwawasa patch where $y>0$ and $\chi\in \mathbb{R}$. 
As is well-known, this is not a global coordinate system of AdS$_2$ and, in particular,  the action of the AdS$_2$ isometry group $SL(2,\mathbb{R})$ does not preserve this patch, unlike for the Euclidean case, see below.
By thinking of the embedding in the ambient space $\mathbb{R}^{1,2}$, 
we can think of all of AdS$_2$ as the above Poincar\'e patch, also allowing values $y<0$. 
This still misses the (light-like in $\mathbb{R}^{1,2}$) hyperplane $y=0$ where the metric 
becomes singular, but this will be of no relevance in our discussion. 

The spectral problem of $\Omega_\alpha$ for $\alpha^2=-2$ is then recognised as being related to the scalar d'Alembertian on AdS$_2$ where it is known that normalisable solutions
\begin{align}
\Omega_\alpha H(\chi,y) = -\mathcal{E} H (\chi,y)
\end{align}
exist for $\mathcal{E}<\frac14$ by the Breitenlohner--Freedman bound~\cite{Breitenlohner:1982jf}. This is the other side of the bound for real roots $\alpha$.
Since the space associated with $\alpha$ now is Lorentzian, the orthogonal space is Euclidean and therefore the corresponding operator $-\Omega_\perp$ becomes elliptic. The remaining equation 
$\Omega_\perp F = \mathcal{E} F$ then has oscillating solutions. 
However, in this case it is not clear what variable now plays the role of $\rho$ that is the variable `towards the singularity' as the geometric interpretation of the solutions involving purely imaginary roots is already unclear at the classical level~\cite{Kleinschmidt:2005gz}. This is due to the lack of a dictionary beyond level $\ell=3$.

There are also mathematical subtleties, related to the ones already 
discussed in \cite{Moore}, that cast doubt on the existence of a 
proper automorphic theory for this case. Namely, 
the action of $SL(2,\mathbb{R})$ in the coordinate system (\ref{AdS})
can be worked out from the ambient space and gives
\begin{align*}
\begin{pmatrix} a & b \\ c&d \end{pmatrix} \cdot y = \frac{y}{(c\chi+d)^2 -(cy)^2} \,,\quad
\begin{pmatrix} a & b \\ c&d \end{pmatrix} \cdot \chi = \frac{ac (\chi^2 - y^2) +(ad +bc) \chi +bd}{(c\chi+d)^2 -(cy)^2}\,,
\end{align*}
This formula resembles the one for M\"obius transformations, except for a `Wick rotation' 
of the $y$ variable, as a consequence of which the coordinate range $y>0$ 
is no longer preserved. 
Independently of the non-preservation of the Poincar\'e patch, we see that
for the generators $T$ and $S$ of the discrete subgroup SL(2,$\mathbb{Z}$), 
this formula implies
\bea
T\, &=&\, \begin{pmatrix} 1 & 1 \\ 0 & 1 \end{pmatrix} \quad : \qquad
(\chi,y) \;\ra\; (\chi+1,y) 
\nn\\[2mm]
S\, &=&\, \begin{pmatrix} 0 & 1 \\ -1 & 0 \end{pmatrix} \quad : \qquad
\chi \pm y \;\ra \; - \frac1{\chi \pm y}
\eea
Hence $T$ and $S$ act as in the Euclidean case, but separately
on the {\em real} null coordinates $\chi\pm y$, as already noted
in \cite{Moore}. As explained there, one can compactify the space by considering $\chi\pm y\in \mathbb{R}\cup \{\infty\}$ so that the space has the topology a two-torus $S^1\times S^1$. Now it is known that the action of SL(2,$\mathbb{Z}$) on the real axis with 
the point at infinity added is dense (any rational number can be mapped to any other
by means of a discrete M\"obius transformation). Because the action on the two defining 
circles is dense there is therefore no sensible fundamental region, unlike for
real roots. We are not aware of a discussion of the consequences of this fact
for the theory of automorphic forms on such a space, nor its implications for the
proper definition of the hypothetical discrete duality group $\E(\mathbb{Z})$.

\section{\texorpdfstring{General comments on $\E$  wave function}{General comments on E10 wave function}}
\label{sec:cmts}

In the previous section we have presented several examples of truncations
of the $\E$ WDW equation to {\em finitely} many variables. In this 
final section we want to return to the general case and collect 
some more general statements. More specifically, the $\E$ wave function 
$\Phi$ being part of a functional representation of $\E$, one can ask the question 
what type of representation component it belongs to if it solves $\Omega\, \Phi =0$. 
As we will see, all indications point towards the necessity of an 
enlargement of the $\E$ framework.

The first observation is that if $\Phi$ belonged to an irreducible highest or lowest weight 
representation of $\E$, then $\Omega \Phi =0$ would imply that 
$\Phi = 1$ is the trivial representation (a statement
that is, of course, familiar from standard group and representation theory).
This is actually in agreement with our findings for the finite-dimensional truncations
studied in the previous sections, as already for the simplest case of 
$\mathfrak{sl}_2$ the relevant eigenfunctions belong to unitary
representations which are neither of highest or lowest weight type.

More generally, this can be seen by recalling that  for such representations we have~\cite{Kac}
\beq 
\Omega \, \Phi = \frac12 \Big( (\L|\L) + 2(\varpi | \L) \Big)\Phi
\eeq
where $\L$ is the relevant highest or lowest weight, and $\varpi$ the Weyl vector.
Now for the fundamental $\lae_{10}$ weights $\L_i$, which obey $(\L_i | \a_j) = \d_{ij}$
we have (see e.g.~\cite{KacMoodyWakimoto})
\beq
(\L_i | \L_j) \leq 0
\eeq
with equality if and only if $\L_i = \L_j = \L_1 = -\d$, the fundamental weight of the `hyperbolic' node. 
Furthermore, for any non-trivial weight $\L = \sum_j p^j \L_j$ we have $(\L|\varpi) < 0$ and thus
\beq
(\L|\L) + 2(\varpi | \L) \,<\, 0
\eeq
This argument shows that for any non-trivial such representation we have 
$\Omega \, \Phi \neq 0$, hence the WDW equation cannot be satisfied.
This conclusion is also in accord with indefiniteness of WDW operator (which here appears
with peculiar and unique ordering prescribed by (\ref{WDW2})): highest (or lowest) weight 
representations are unitarisable \cite{Kac}, whereas for standard WDW equation we
have the usual indefinite metric Hilbert space, just like for the Klein--Gordon  
wave function. This again leads to the conclusion that $\Phi$ cannot belong to 
a highest or lowest weight representation of $\E$.

\medskip

In the foregoing section, we have considered differential operators that only depend on the $\E$ coordinates up to $\mathfrak{gl}_{10}$ level $\ell\leq 2$. Such a truncation breaks $E_{10}$ symmetry, but it is possible to solve  the equation $\Omega\, \Phi=0$ in such a truncation consistently. This statement is analogous to the statement for the classical $\E$ coset model that one can truncate the geodesic equation such that only finitely many coset velocity components are non-zero but this provides a solution to the full geodesic equation~\cite{Cocoyoc}. At the level of induced representations and automorphic forms it corresponds to considering restricted Fourier coefficients, i.e., to perform the Fourier integral over all variables of $\ell>2$~\cite{Fleig:2015vky}.

Although such truncations are thus all consistent, it is another question whether they
are also stable in the full configuration space w.r.t. small perturbations along
the truncated directions. A relevant fact here is that, as shown in \cite{DN}, 
{\em classical} geodesics on the $\E/\K$ coset manifold are infinitely unstable. 
We recall that the geodesic deviation equations governing the relative evolution
of two neighbouring geodesics are determined by the sectional curvatures
(see {\em e.g.} \cite{Weinberg}). Consequently, for a geodesic with tangent vector 
$v\in\lah$ in the Cartan subalgebra, and a  deviation in the direction of the generator
$E_\a^+ \equiv E_\a + F_\a$ for an arbitrary root $\a$, the deviation 
of the two geodesics is determined by (see appendix of \cite{DN})
\beq\label{GD}
\cR(v,E^+_\a,v,E^+_\a) \,=\, - \big( \a(v) \big)^2 \,<\,  0
\eeq
This expression decreases without bound for imaginary roots.
This is because for every imaginary root $\a$, any integer multiple $n\a$ is also 
a root. Hence, replacing $\a$ by $n\a$ on the r.h.s., (\ref{GD}) can be made arbitrarily 
negative by taking $n\ra\infty$, with an exponentially increasing number
of unstable directions (labeled by the multiplicity index of the root $n\a$)  
for time-like imaginary roots. It is not clear how this instability is reflected 
in the $\E$ WDW operator, although the usual formal path 
integral representation of the `wave function of the universe' (see {\em e.g.}
\cite{Halliwell}) would 
suggest that the instability should manifest itself via the saddle point approximation.
It is also unclear how the inclusion of fermions
and third quantisation might affect these conclusions.

Finally, we should point out that the picture here with a
wave function that vanishes at the singularity is very different from the one 
suggested by the no boundary proposal of \cite{Hartle:1983ai} (see also \cite{Nic} and 
\cite{Perry:2021mch} for related discussions). The latter hypothesises a creation of the universe `out of nothing' 
in terms of a Euclidean instanton, where in particular the BKL analysis and chaotic oscillations play no role. The absence of an initial singularity hinges on the presence
of a  {\em positive} cosmological constant (which is known to suppress chaotic 
oscillations \cite{Damour:2002et}). By contrast,  $D=11$ supergravity does not admit 
a non-vanishing cosmological constant. While a cosmological constant is almost
always generated by spontaneous compactification, it generically turns out 
to be negative. By contrast, the (classical)  $\E/\K$ model has been shown to 
not admit static solutions, but rather gives rise to a time-dependent cosmological 
evolution of quintessence  type \cite{Kleinschmidt:2005gz}.

\subsubsection*{Acknowledgements}

We are grateful to Guillaume Bossard, Klaus Fredenhagen, Marc Henneaux, 
Ralf K\"ohl, Robin Lautenbacher, Jean-Luc Lehners, Victor Lekeu, Gregory Moore and Stefan Theisen
for discussions and correspondence related to this work. The work of H.N. has received funding 
from the European Research Council (ERC) under the European Union's Horizon 2020 
research and innovation programme (grant agreement No 740209).

\appendix

\section{Theory with both three-form and six-form}
\label{app:HT}

In the covariant theory~\eqref{eq:SUGRA11}, the matter equation of motion 
\begin{align}
\partial_S (E F^{SMNP}) = -\frac1{576} \ve^{MNPK_1\ldots K_8} F_{K_1\ldots K_4} F_{K_5\ldots K_8}
\end{align}
allows for the introduction of a dual seven-form field strength according to
\begin{align}
\label{eq:dual11}
F_{M_1\ldots M_7} =  \frac{1}{4!} \epsilon_{M_1\ldots M_7 N_1\ldots N_4} F^{N_1\ldots N_4}
\quad\Leftrightarrow\quad  F_{M_1\ldots M_4} =  - \frac1{7!} \epsilon_{M_1\ldots M_4 N_1\ldots N_7} F^{N_1\ldots N_7} \,,
\end{align}
where the seven-form field strength is given by
\begin{align}
\label{eq:defF7}
F_{M_1\ldots M_7} = 7 \partial_{[M_1} A_{M_2\ldots M_7}] -35  A_{[M_1 M_2 M_3} F_{M_4\ldots M_7]}
\end{align}
and satisfies the modified Bianchi identity
\begin{align}\label{nonabBianchi}
8 \partial_{[M_1} F_{M_2\ldots M_8]} = -70 F_{[M_1\ldots M_4} F_{M_5\ldots M_8]}
\end{align}
As usual, the duality relation exchanges Bianchi identities and equations of motion and the extra term in the definition of the seven-form field strength is chosen such that duality is compatible with the three-form equation of motion above.

It was shown in~\cite{Henneaux:1988gg,Bandos:1997gd,Bunster:2011qp} that such a six-form potential can already be introduced at the level of the action by breaking manifest space-time covariance.\footnote{Writing the non-linear theory solely in terms of the six-form is not possible~\cite{Nicolai:1980kb}.} As a six-form potential appears in the $\E$ theory (which does not exhibit manifest Lorentz symmetry either), we now switch to this formulation, following~\cite{Bunster:2011qp}. We focus solely on the matter sector and will leave the gravitational sector untouched in this appendix.

The first step is to explicitly solve the Gauss constraint~\eqref{eq:Gauss} in terms of the differential of a dual six-form\footnote{Here, we work locally and thus there are no topological obstructions to this application of the Poincar\'e lemma.} 
\begin{align}
\label{eq:GaussSol}
\Pi^{mnp} +\frac1{3\cdot 144} \ve^{mnpk_1\ldots k_7} A_{k_1k_2k_3} F_{k_4\ldots k_7} =  \frac{1}{6!} \ve^{mnp k_1\ldots k_7}   \partial_{k_1} A_{k_2\ldots k_7}\,,
\end{align}
which is similar to the duality relation~\eqref{eq:defF7}.
Inserting this solution leads to the canonical action 
\begin{align}
\label{eq:2pot}
\mathcal{L}_{\text{can}} &\,=\, \frac12 \dot{g}_{mn} \Pi^{mn}  + \frac1{3!} \dot{A}_{mnp}  \ve^{mnpk_1\ldots k_7}\left(\frac{1}{6!} \partial_{k_1} A_{k_2\ldots k_7} - \frac1{3\cdot 144}A_{k_1k_2k_3} F_{k_4\ldots k_7} \right) \nn\\
&\quad\quad - N \mathcal{H} -N^m \mathcal{H}_m 
\end{align}
that depends only on the spatial components of the three-form $A_{mnp}$ and its dual six-form $A_{m_1\ldots m_6}$. The matter Hamiltonian $e\mathcal{H}^{\text{(mat)}}$ from~\eqref{eq:H11} can now be written as
\begin{align}
&\quad \frac1{12} \big(\Pi^{mnp} - \mathcal{P}^{mnp} \big) g_{mm'}g_{nn'}g_{pp'}
 \big(\Pi^{m'n'p'} - \mathcal{P}^{m'n'p'} \big) +\frac1{48}e^2 F_{m_1\ldots m_4}g^{m_1n_1}\cdots g^{m_4n_4}F_{n_1\ldots n_4}\nn\\
 &= \frac{1}{2\cdot 7!}e^2 F_{m_1\ldots m_7} g^{m_1n_1}\cdots g^{m_7n_7} F_{n_1\ldots n_7} +\frac1{2\cdot 4!} e^2 F_{m_1\ldots m_4}g^{m_1n_1}\cdots g^{m_4n_4}F_{n_1\ldots n_4}\,,
\end{align}
where 
\begin{align}
F_{m_1\ldots m_7} &= -\frac1{3!} \ve_{m_1\ldots m_7 n_1n_2n_3} \left( \Pi^{n_1n_2n_3} - \mathcal{P}^{n_1n_2n_3}\right) \nn\\
&= 7\partial_{[m_1} A_{m_2\ldots m_7]}  -35 A_{[m_1m_2m_3} F_{m_4\ldots m_7]}\,,
\end{align}
and where we used~\eqref{eq:GaussSol} on the solution of the Gauss constraint, an answer that is consistent with the spatial components of~\eqref{eq:defF7}. Note that $F_{m_1\ldots m_7}$ is tensorial.

The variation of the matter part of the action~\eqref{eq:2pot} with respect to $A_{mnp}$ and  $A_{m_1\ldots m_6}$ gives the equations of motion 
\begin{subequations}
\begin{align}
\label{eq:1}
0&= - \frac{1}{3!\cdot 6!}  \ve^{mnpk_1\ldots k_7} \partial_{k_1} \left(\partial_t A_{k_2\ldots k_7}  -20A_{k_2k_3k_4}  \partial_t A_{k_5k_6k_7} +\frac{E}{4!} \ve_{k_2\ldots k_7 n_1\ldots n_4} F^{n_1\ldots n_4}\right) \\
&\quad + \frac{1}{216} \ve^{mnpk_1\ldots k_7} A_{k_1k_2k_3} \partial_{k_4} \partial_t A_{k_5k_6k_7} 
 + \frac{E}{72}  F^{mnp k_1\ldots k_4} F_{k_1\ldots k_4}\nn\\
 &\quad\quad  - \frac1{36} \partial_s(E F^{smnp k_1k_2k_3} ) A_{k_1k_2k_3} \,, \nn \\[2mm]
 \label{eq:2}
0&= \partial_m \left( \frac{1}{3!} \ve^{k_1\ldots k_6 m n_1\ldots n_3} \dot{A}_{n_1n_2n_3} + E F^{mk_1\ldots k_6} \right) \,.
\end{align}
\end{subequations}
We reiterate that we focus on the matter sector only here and we also work in flat space-time for simplicity.
The second equation is solved locally by
\begin{align}
\label{eq:2sol}
E F^{mk_1\ldots k_6}  +  \frac{1}{3!} \ve^{k_1\ldots k_6 m n_1\ldots n_3} \dot{A}_{n_1n_2n_3} =  \frac12  \ve^{k_1\ldots k_6mn_1n_2n_3} \partial_{n_1} A_{tn_2n_3} 
\end{align}
for some function $A_{tn_2n_3}$. Rewriting this formula leads to
\begin{align}
E F^{k_1\ldots k_7}  = - \frac1{3!} \ve^{k_1\ldots k_7 n_1n_2n_3} F_{tn_1n_2n_3}\quad \Leftrightarrow\quad
F_{tn_1n_2n_3} = -  \frac{1}{7!}  \epsilon_{tn_1n_2n_3k_1\ldots k_7 }  F^{k_1\ldots k_7} \,,
\end{align}
where we have reintroduced the time index on the Levi--Civita symbol and turned it into the Levi--Civita tensor by absorbing $E$ in order to recognise this equation as the time component of the second way of writing the duality equation~\eqref{eq:dual11}.

Using~\eqref{eq:2} and its solution~\eqref{eq:2sol} in the first equation of motion~\eqref{eq:1} we get an exterior derivative
\begin{align}
0&= - \frac{1}{3!\cdot 6!}  \ve^{mnpk_1\ldots k_7} \partial_{k_1} \left(\partial_t A_{k_2\ldots k_7} -35 A_{[tk_2k_3} \partial_{k_4} A_{k_5k_6k_7]}+\frac{E}{4!} \ve_{k_2\ldots k_7 n_1\ldots n_4} F^{n_1\ldots n_4}\right) \,,
\end{align}
where in particular the term with the bare $A_{tk_2k_3}$ inside the derivative comes from using~\eqref{eq:2sol}. The above equation can be solved locally by 
\begin{align}
\partial_t A_{k_2\ldots k_7} -35 A_{[tk_2k_3} \partial_{k_4} A_{k_5k_6k_7]}+\frac{E}{4!} \ve_{k_2\ldots k_7 n_1\ldots n_4} F^{n_1\ldots n_4} = - 6\partial_{[k_2} A_{k_3\ldots k_7]t}
\end{align}
introducing a function that plays the role of the time component of the six-form potential. Rewriting the equation we then find
\begin{align}
F_{tn_1\ldots n_6} = \frac1{4!} \epsilon_{tn_1\ldots n_6 k_1\ldots k_4} F^{k_1\ldots k_4}\,,
\end{align}
where we have introduced the time index and turned the Levi--Civita symbol into its tensor form. This agrees perfectly with the time component of the first way of writing the duality equation~\eqref{eq:dual11}.

The kinetic term can be brought into a slightly more symmetric form by using integration by parts
\begin{align}
&\quad \frac1{3!} \dot{A}_{mnp}  \ve^{mnpk_1\ldots k_7}\left(\frac{1}{6!} \partial_{k_1} A_{k_2\ldots k_7} - \frac1{3\cdot 144}A_{k_1k_2k_3} F_{k_4\ldots k_7} \right)\nn\\
&= \frac{1}{2\cdot 3!\cdot 7!} \dot{A}_{mnp}\ve^{mnpk_1\ldots k_7} F_{k_1\ldots k_7} - \frac{1}{2\cdot 4! \cdot6!} F_{mnpq} \ve^{mnpqk_1\ldots k_6} \dot{A}_{k_1\ldots k_6}
\nn\\
&\quad 
+ \frac{1}{3!\cdot 3!\cdot 144} \dot{A}_{mnp} \ve^{mnpk_1\ldots k_7} A_{k_1k_2k_3} F_{k_4\ldots k_7}
\end{align}
In this form both fields appear with time derivatives and have non-vanishing canonical momenta. 

Starting then from the action 
\begin{align}
\mathcal{L} &=  \frac{1}{2\cdot 3!\cdot 7!} \dot{A}_{mnp}\ve^{mnpk_1\ldots k_7} F_{k_1\ldots k_7} - \frac{1}{2\cdot 4! \cdot6!} F_{mnpq} \ve^{mnpqk_1\ldots k_6} \dot{A}_{k_1\ldots k_6}
\nn\\
&\quad 
+ \frac{1}{3!\cdot 3!\cdot 144} \dot{A}_{mnp} \ve^{mnpk_1\ldots k_7} A_{k_1k_2k_3} F_{k_4\ldots k_7}
\\
&\quad
- N \left[ \frac{1}{2\cdot 7!}e F_{m_1\ldots m_7} g^{m_1n_1}\cdots g^{m_7n_7} F_{n_1\ldots n_7} +\frac1{2\cdot 4!} e F_{m_1\ldots m_4}g^{m_1n_1}\cdots g^{m_4n_4}F_{n_1\ldots n_4} \right]\nn
\end{align}
we get the canonical momenta
\begin{align}
\Pi^{mnp} &= \frac1{2\cdot 7!} \ve^{mnpk_1\ldots k_7} F_{k_1\ldots k_7} + \frac1{3!\cdot 144} \ve^{mnpk_1\ldots k_7} A_{k_1k_2k_3} F_{k_4\ldots k_7}\,\nn\\
\Pi^{m_1\ldots m_6} &= - \frac1{2\cdot 4!} \ve^{m_1\ldots m_6 k_1\ldots k_4} F_{k_1\ldots k_4}\,.
\end{align}
They represent primary constraints of the theory and are analysed in detail in section~\ref{sec:11D}.

\section{\texorpdfstring{More details on the $\E$ Casimir}{More details on the E10 Casimir}}
\label{app:Cas}

The $\E$ Casimir with parabolic normal ordering was given in~\eqref{WDW2}. The terms involving the 
GL(10) generators $K^m{}_n$ were not fully normal-ordered there, and therefore the coefficient
of the linear term is not the same as for the standard expressions for the Casimir operator.
Up to normalisation, when acting on integrable highest weight modules, the latter is generally 
given by the fully normal ordered expression \cite{Kac} 
\beq\label{Casimir}
\Omega \,=\, \frac12 G^{ab} H_a H_b \, +\,  G^{ab} \varpi_a H_b \,+\,  
\sum_{\a >0} \sum_{s=1}^{{\rm mult}(\a)} E_{-\a}^s E_\a^s
\eeq
where $\varpi$ is the Weyl vector, and the sum on the r.h.s. runs over
all positive roots together with their multiplicities. In this appendix we show that 
for $\E$ the two expressions \eqref{WDW2} and (\ref{Casimir}) are, in fact, the same.

Normal-ordering the GL(10) terms in~\eqref{WDW2} yields
\begin{align}
&\quad \frac12 K^m{}_n K^n{}_m -\frac1{18} KK +\frac{23}{6} K \\
&= \sum_{m>n} K^m{}_n K^n{}_m + \frac12 \sum_{m<n} \left[K^m{}_n ,K^n{}_m\right] + \frac12\sum_m K^m{}_m K^m{}_m - \frac1{18} K K + \frac{23}{6} K\nn\\
&=  \sum_{m>n} K^m{}_n K^n{}_m + \frac12 \sum_{m<n} (K^m{}_m -K^n{}_n) + \frac12\sum_m K^m{}_m K^m{}_m - \frac1{18} K K + \frac{23}{6} K     & \nn
\end{align}
Now we use
\bea
\frac12 \sum_{m<n} (K^m{}_m - K^n{}_n) + \frac{23}6 K &=&
 \frac12 \big( 9 K^1{}_1 + 7 K^2{}_2 \cdots - 9K^{10}{}_{10} \big) +\frac{23}6 K   \nn\\[1mm]
&=&
 \frac13 \big( 25 K^1{}_1 + 22 K^2{}_2 + \cdots + K^9{}_9 - 2 K^{10}{}_{10} \big)\nn\\
 &=&G^{ab}\varpi_a H_b
\eea
where in the last expression we have used the $\E$ Weyl vector in the wall basis
\beq
\varpi \,=\, (-30, -31, \cdots, - 39) \,.
\eeq
together with \cite{Cocoyoc}
\begin{align}
H_1 = K^2{}_2 -K^1{}_1\; , \cdots , \, H_9 = K^{10}{}_{10} - K^{9}{}_{9} \;,\;
H_{10} =  K^8{}_8 + K^9{}_9 + K^{10}{}_{10} -  \frac13 K
\end{align}
The quadratic terms are as they should be for the fully normal ordered 
$\E$ Casimir. In conclusion, after normal ordering the GL(10) contribution, 
our expressions for the Casimir precisely coincide.

The coefficient $\frac{23}{6}$ multiplying $K$ in~\eqref{WDW2} can be generally understood as follows. We explain the calculation for any $E_D$ decomposed with respect to its obvious GL$(D)$ subgroup. The linear term arises from normal-ordering all terms with roots on GL$(D)$ levels $\ell>0$, therefore it equals
\begin{align}
\beta \equiv \frac 12 \sum_{\substack{\alpha> 0\\\ell>0}} \alpha = \frac12 \sum_{\alpha>0} \alpha - \frac 12 \sum_{\substack{\alpha> 0\\\ell=0}} \alpha = \varpi_{E_{D}} -\frac 12 \sum_{\substack{\alpha> 0\\\ell=0}} \alpha 
\end{align}
For Kac--Moody $E_D$, the sums are divergent over infinitely many positive roots are ill-defined but the Weyl vector is well-defined, so $\beta$ is a well-defined element. The sum over the positive roots on level $\ell=0$ gives the $GL(D)$ Weyl vector whose form in a simple root basis is
\begin{align}
\frac 12 \sum_{\substack{\alpha> 0\\\ell=0}} \alpha  = 
\frac12 \big( D-1\,,\, 2(D-2)\,,\, 3(D-3)\,,\, \ldots, 2(D-2)\,,\, D-1\,,\,0\big)\,.
\end{align}
The inner product of $\beta$ with all simple roots can be computed as
\begin{align}
\beta \cdot \alpha_D = 1+\frac12 3(D-3) = \frac{3D-7}{2}\,,\quad\quad \beta \cdot \alpha_i = 0 \quad\text{for $i\neq D$}
\end{align}
since the exceptional node attaches three nodes from the end of the $GL(D)$ line. These inner products identify $\beta=\frac{3D-7}{2}\Lambda_D$ in the basis of fundamental weights. Using moreover that $K = 3\Lambda_D^\vee$ we deduce that the linear term is $\frac{3D-7}{6}K$ that for $D=10$ gives the claimed value.


\baselineskip15pt

\begin{thebibliography}{90}

\bibitem{DeWitt} B.~S.~DeWitt, 
``Quantum Theory of Gravity. 1. The Canonical Theory,''
\doi{Phys. Rev. \textbf{160} (1967), 1113-1148}{doi:10.1103/PhysRev.160.1113};
``Quantum Theory of Gravity. 2. The Manifestly Covariant Theory,''
\doi{Phys. Rev. \textbf{162} (1967), 1195-1239}{doi:10.1103/PhysRev.162.1195}.

\bibitem{Wheeler} J.A. Wheeler,
``Superspace and the nature of quantum geometrodynamics",
Adv. Ser. Astrophys. Cosmol. {\bf 3} (1968) 27.

\bibitem{Kiefer:2007ria}
C.~Kiefer,
\doi{{\sl Quantum Gravity}}{DOI:10.1093/acprof:oso/9780199585205.001.0001}, 3rd edition, 
Oxford University Press (Oxford, 2012).

\bibitem{Carlip} S. Carlip,
\doi{{\sl Quantum Gravity in $2+1$ dimensions}}{doi:10.1017/CBO9780511564192},
Cambridge Monographs on Mathematical Physics, 
Cambridge University Press (1998).

\bibitem{Damour:2002cu}
T.~Damour, M.~Henneaux and H.~Nicolai,
``E(10) and a 'small tension expansion' of M theory,''
\doi{Phys. Rev. Lett. \textbf{89} (2002), 221601}{doi:10.1103/PhysRevLett.89.221601}
\eprint{hep-th/0207267}.

\bibitem{Cocoyoc} T.~Damour and H.~Nicolai,
``Eleven dimensional supergravity and the $\E/\K$ sigma-model at low A(9) levels,''
in: Group Theoretical Methods in Physics, Institute of
  Physics Conference Series No. 185, IoP Publishing, 2005
\eprint{hep-th/0410245}.

\bibitem{KN0} A. Kleinschmidt and H. Nicolai,
``Maximal supergravities and the E(10) coset model",
\doi{Int. J. Mod. Phys. D \textbf{15} (2006), 1619-1642}{doi:10.1142/S0218271806009005}.

\bibitem{Ju85} B.~Julia,
 ``Kac-Moody symmetry of gravitation and supergravity theories,''
 in: Lectures in Applied Mathematics, Vol. 21
  (1985), AMS-SIAM, p. 335.

\bibitem{Cremmer:1978km}
E.~Cremmer, B.~Julia and J.~Scherk,
``Supergravity Theory in Eleven-Dimensions,''
\doi{Phys. Lett. B \textbf{76} (1978), 409-412}{doi:10.1016/0370-2693(78)90894-8}.

\bibitem{Diaz:1986jw}
A.~H.~Diaz,
``Hamiltonian formulation of eleven-dimensional supergravity,''
\doi{Phys. Rev. D \textbf{33} (1986), 2801-2808}{doi:10.1103/PhysRevD.33.2801}.

\bibitem{Diaz:1986jx}
A.~H.~Diaz,
``Constraint algebra in eleven-dimensional supergravity,''
\doi{Phys. Rev. D \textbf{33} (1986), 2809-2812}{doi:10.1103/PhysRevD.33.2809}.

\bibitem{Bunster:2011qp}
C.~Bunster and M.~Henneaux,
``The Action for Twisted Self-Duality,''
\doi{Phys. Rev. D \textbf{83} (2011), 125015}{doi:10.1103/PhysRevD.83.125015}
\eprintN{1103.3621}.

\bibitem{Bandos:1997gd}
I.~A.~Bandos, N.~Berkovits and D.~P.~Sorokin,
``Duality symmetric eleven-dimensional supergravity and its coupling to M-branes,''
\doi{Nucl. Phys. B \textbf{522} (1998), 214-233}{doi:10.1016/S0550-3213(98)00102-3}
\eprint{hep-th/9711055}.

\bibitem{West:2001as}
P.~C.~West,
``E(11) and M theory,''
\doi{Class. Quant. Grav. \textbf{18} (2001), 4443-4460}{doi:10.1088/0264-9381/18/21/305}
\eprint{hep-th/0104081}.

\bibitem{Glennon:2020uov}
K.~Glennon and P.~West,
``The non-linear dual gravity equation of motion in eleven dimensions,''
\doi{Phys. Lett. B \textbf{809}, 135714 (2020)}{doi:10.1016/j.physletb.2020.135714}
\eprintN{2006.02383}.

\bibitem{Belinsky:1970ew}
V.~A.~Belinsky, I.~M.~Khalatnikov and E.~M.~Lifshitz,
``Oscillatory approach to a singular point in the relativistic cosmology,''
\doi{Adv. Phys. \textbf{19} (1970), 525-573}{doi:10.1080/00018737000101171}.

\bibitem{Hohm:2013jma}
O.~Hohm and H.~Samtleben,
``U-duality covariant gravity,''
\doi{JHEP \textbf{09} (2013), 080}{doi:10.1007/JHEP09(2013)080}
\eprintN{1307.0509}.

\bibitem{Hohm:2013pua}
O.~Hohm and H.~Samtleben,
``Exceptional Form of D=11 Supergravity,''
\doi{Phys. Rev. Lett. \textbf{111} (2013), 231601}{doi:10.1103/PhysRevLett.111.231601}
\eprintN{1308.1673}.
  
\bibitem{BKS}
 G. Bossard, A. Kleinschmidt and E. Sezgin,
``A master exceptional field theory",
\doi{JHEP \textbf{06} (2021), 185}{doi:10.1007/JHEP06(2021)185}
\eprintN{2103.13411}.
  
 \bibitem{Damour:2006xu}
T.~Damour, A.~Kleinschmidt and H.~Nicolai,
``K(E(10)), Supergravity and Fermions,''
\doi{JHEP \textbf{08} (2006), 046}{doi:10.1088/1126-6708/2006/08/046}
\eprint{hep-th/0606105}.

\bibitem{Kleinschmidt:2014uwa}
A.~Kleinschmidt, H.~Nicolai and N.~K.~Chidambaram,
``Canonical structure of the E10 model and supersymmetry,''
\doi{Phys. Rev. D \textbf{91} (2015) no.8, 085039}{doi:10.1103/PhysRevD.91.085039}
\eprintN{1411.5893}.
 
\bibitem{DKN} T. Damour, A. Kleinschmidt and H. Nicolai,
``Constraints and the E10 coset model",
\doi{Class. Quant. Grav. \textbf{24} (2007), 6097-6120}{doi:10.1088/0264-9381/24/23/025}
\eprintN{0709.2691}.

\bibitem{Damour:2009ww}
T.~Damour, A.~Kleinschmidt and H.~Nicolai,
``Sugawara-type constraints in hyperbolic coset models,''
\doi{Commun. Math. Phys. \textbf{302} (2011), 755-788}{doi:10.1007/s00220-011-1188-y}
\eprintN{0912.3491}.

\bibitem{Bossard:2017wxl}
G.~Bossard, A.~Kleinschmidt, J.~Palmkvist, C.~N.~Pope and E.~Sezgin,
``Beyond E$_{11}$,''
\doi{JHEP \textbf{05} (2017), 020}{doi:10.1007/JHEP05(2017)020}
\eprintN{1703.01305}.

\bibitem{Cederwall:2021ymp}
M.~Cederwall and J.~Palmkvist,
``Tensor hierarchy algebra extensions of over-extend\-ed Kac--Moody algebras,''
\doi{Commun. Math. Phys. (2021)}{doi:10.1007/s00220-021-04243-3}
\eprintRT{2103.02476}.

\bibitem{Hull:2001iu}
C.~M.~Hull,
``Duality in gravity and higher spin gauge fields,''
\doi{JHEP \textbf{09} (2001), 027}{doi:10.1088/1126-6708/2001/09/027}
\eprint{hep-th/0107149}.

\bibitem{Bergshoeff:2008vc}
E.~A.~Bergshoeff, M.~de Roo, S.~F.~Kerstan, A.~Kleinschmidt and F.~Riccioni,
``Dual Gravity and Matter,''
\doi{Gen. Rel. Grav. \textbf{41} (2009), 39-48}{doi:10.1007/s10714-008-0650-4}
\eprintN{0803.1963}.

\bibitem{Fleig:2015vky}
  P.~Fleig, H.~P.~A.~Gustafsson, A.~Kleinschmidt and D.~Persson,
  \doi{{\em Eisenstein series and automorphic representations. With applications in string theory}}{https://doi.org/10.1017/9781316995860}, Cambridge University Press (2018). Preliminary version at  \eprintNT{1511.04265}.

\bibitem{Hull:1994ys}
C.~M.~Hull and P.~K.~Townsend,
``Unity of superstring dualities,''
\doi{Nucl. Phys. B \textbf{438} (1995), 109-137}{doi:10.1016/0550-3213(94)00559-W}
\eprint{hep-th/9410167}.

\bibitem{Ganor:1999ui}
O.~J.~Ganor,
``Two conjectures on gauge theories, gravity, and infinite dimensional Kac--Moody groups,''
\eprint{hep-th/9903110}.

\bibitem{Brown:2004jb}
J.~Brown, O.~J.~Ganor and C.~Helfgott,
``M theory and E(10): Billiards, branes, and imaginary roots,''
\doi{JHEP \textbf{08} (2004), 063}{doi:10.1088/1126-6708/2004/08/063}
\eprint{hep-th/0401053}.

\bibitem{Fleig:2012xa}
P.~Fleig and A.~Kleinschmidt,
``Eisenstein series for infinite-dimensional U-duality groups,''
\doi{JHEP \textbf{06} (2012), 054}{doi:10.1007/JHEP06(2012)054}
\eprintN{1204.3043}.

\bibitem{Kumar:2002}
S.~Kumar, \doi{{\em {Kac-Moody
  Groups, their Flag Varieties and Representation Theory}}}{http://dx.doi.org/10.1007/978-1-4612-0105-2}.
\newblock Progress in Mathematics. Birkh\"auser Basel, 2002.

\bibitem{Marquis:2018}
 T.~Marquis,
 \doi{{\sl An introduction to Kac-Moody groups over fields}}{10.4171/187}, 
  EMS Textbooks in Mathematics (European Mathematical Society, 2018).

  \bibitem{KacPeterson}
  V.~G.~Kac and D.~H.~Peterson,
  ``Defining relations of certain infinite dimensional groups,''
  {\hypersetup{urlcolor=darkred}\href{http://www.numdam.org/item/AST_1985__S131__165_0}{Ast\'erisque Hors-S\'erie (1985), 165}\hypersetup{urlcolor=blue}}.

\bibitem{Kleinschmidt:2009cv}
A.~Kleinschmidt, M.~Koehn and H.~Nicolai,
``Supersymmetric quantum cosmological billiards,''
\doi{Phys. Rev. D \textbf{80} (2009), 061701}{doi:10.1103/PhysRevD.80.061701}
\eprintGR{0907.3048}.

\bibitem{Kleinschmidt:2009hv}
A.~Kleinschmidt and H.~Nicolai,
``Cosmological quantum billiards,'' in: \textit{Foundations of Space and Time: Reflections on Quantum Gravity}, J.~Murugan, A.~Weltman and G.~.F.~R.~Ellis (eds.), Cambridge Univ. Press (2002) 106--124
\eprintGR{0912.0854}.

\bibitem{Hartle:1983ai}
J.~B.~Hartle and S.~W.~Hawking,
``Wave Function of the Universe,''
\doi{Phys. Rev. D \textbf{28}, 2960-2975 (1983)}{doi:10.1103/PhysRevD.28.2960}.

\bibitem{Perry:2021mch}
M.~J.~Perry,
``No Future in Black Holes,''
\eprintN{2106.03715}.

\bibitem{Nic} H. Nicolai
``Complexity and the Big Bang",
\doi{Class. Quant. Grav. {\bf 38} (2021) 18, 187001}{doi:10.1088/1361-6382/ac1b07}
\eprintGR{2104.09626}.

\bibitem{DN} T.~Damour and H. Nicolai,
``Higher order M theory corrections and the Kac-Moody algebra $\E$",
\doi{Class. Quant. Grav. \textbf{22} (2005), 2849-2880}{doi:10.1088/0264-9381/22/14/003}
\eprint{hep-th/0504153}.


\bibitem{deBuyl:2005zy}
S.~de Buyl, M.~Henneaux and L.~Paulot,
``Hidden symmetries and Dirac fermions,''
\doi{Class. Quant. Grav. \textbf{22} (2005), 3595-3622}{doi:10.1088/0264-9381/22/17/018}
\eprint{hep-th/0506009}.

\bibitem{Damour:2005zs}
T.~Damour, A.~Kleinschmidt and H.~Nicolai,
``Hidden symmetries and the fermionic sector of eleven-dimensional supergravity,''
\doi{Phys. Lett. B \textbf{634} (2006), 319-324}{doi:10.1016/j.physletb.2006.01.015}
\eprint{hep-th/0512163}.

\bibitem{deBuyl:2005sch}
S.~de Buyl, M.~Henneaux and L.~Paulot,
``Extended E(8) invariance of 11-dimensional supergravity,''
\doi{JHEP \textbf{02} (2006), 056}{doi:10.1088/1126-6708/2006/02/056}
\eprint{hep-th/0512292}.

\bibitem{Damour:2013eua}
T.~Damour and P.~Spindel,
``Quantum supersymmetric cosmology and its hidden Kac--Moody structure,''
\doi{Class. Quant. Grav. \textbf{30} (2013), 162001}{doi:10.1088/0264-9381/30/16/162001}
\eprintGR{1304.6381}.

\bibitem{Damour:2014cba}
T.~Damour and P.~Spindel,
``Quantum Supersymmetric Bianchi IX Cosmology,''
\doi{Phys. Rev. D \textbf{90} (2014) no.10, 103509}{doi:10.1103/PhysRevD.90.103509}
\eprintGR{1406.1309}.

\bibitem{Damour:2022oja}
T.~Damour and P.~Spindel,
``Hidden Kac-Moody Structures in the Fermionic Sector of Five-Dimensional Supergravity,''
\eprintN{2202.03794}.

\bibitem{Nahm} P. Goddard, W. Nahm and D.I. Olive,
``Symmetric Spaces, Sugawara's Energy Momentum Tensor 
in Two-Dimensions and Free Fermions",
\doi{Phys. Lett. {\bf B160} (1985) 111}{doi:10.1016/0370-2693(85)91475-3}.

\bibitem{Henneaux:1988gg}
M.~Henneaux and C.~Teitelboim,
``Dynamics of Chiral (Selfdual) $p$ Forms,''
\doi{Phys. Lett. B \textbf{206} (1988), 650-654}{doi:10.1016/0370-2693(88)90712-5}.

\bibitem{Henneaux:1987hz}
M.~Henneaux and C.~Teitelboim,
``Consistent quantum mechanics of chiral p forms,''
in: \textit{Quantum mechanics of fundamental systems II}, C.~Teitelboim and J.~Zanelli (eds.), 79--112, Plenum Press (New York, 1989).

\bibitem{MN} H. Nicolai and H.J. Matschull,
``Aspects of canonical gravity and supergravity,''
\doi{J. Geom. Phys. \textbf{11} (1993), 15-62}{doi:10.1016/0393-0440(93)90047-I}.

\bibitem{Henneaux:1986cz}
M.~Henneaux,
``Hamiltonian formulation of D=10 supergravity theories,''
\doi{Phys. Lett. B \textbf{168} (1986), 233-238}{doi:10.1016/0370-2693(86)90970-6}.

\bibitem{Kreutzer:2020quf}
L.~T.~Kreutzer,
``Canonical analysis of E$_{6(6)}(\mathbb{R})$ invariant five dimensional \text{(super-)}
gravity,''
\doi{J. Math. Phys. \textbf{62} (2021) no.3, 032302}{doi:10.1063/5.0037092}
\eprintN{2005.13553}.

\bibitem{Nicolai:1980kb}
H.~Nicolai, P.~K.~Townsend and P.~van Nieuwenhuizen,
``Comments on 11-dimensional supergravity,''
\doi{Lett. Nuovo Cim. \textbf{30} (1981), 315}{doi:10.1007/BF02817085}.

\bibitem{Henneaux:2004jw}
M.~Henneaux and C.~Teitelboim,
``Duality in linearized gravity,''
\doi{Phys. Rev. D \textbf{71} (2005), 024018}{doi:10.1103/PhysRevD.71.024018}
\eprint{gr-qc/0408101}.

\bibitem{Julia:2005ze}
B.~Julia, J.~Levie and S.~Ray,
``Gravitational duality near de Sitter space,''
\doi{JHEP \textbf{11} (2005), 025}{doi:10.1088/1126-6708/2005/11/025}
\eprint{hep-th/0507262}.

\bibitem{Nicolai:2003fw}
  H.~Nicolai and T.~Fischbacher,
  ``Low level representations for E$_{10}$ and E$_{11}$,''
  Contribution to the Proceedings of the Ramanujan International Symposium on Kac--Moody Algebras and Applications, ISKMAA-2002, Chennai, India, 28--31 January, 
  \eprint{hep-th/0301017}.

\bibitem{NS} H.~Nicolai and H. Samtleben,
``Integrability and canonical structure of $d=2, N=16$ supergravity",
\doi{Nucl. Phys. B \textbf{533} (1998), 210--242}{doi:10.1016/S0550-3213(98)00496-9}
\eprint{hep-th/9804152}.

\bibitem{Kac:1968}
  V.~G.~Kac,
``Simple irreducible graded Lie algebras of finite growth,''
Izv. Akad. Nauk SSSR Ser. Mat. 32 1968 1323--1367. 

\bibitem{Kac} V.~G.~Kac, \doi{{\sl Infinite dimensional Lie algebras}}{https://doi.org/10.1017/CBO9780511626234}, 3rd
  edition, Cambridge University Press (Cambridge, 1990).
  
\bibitem{Bossard:2019ksx}
G.~Bossard, A.~Kleinschmidt and E.~Sezgin,
``On supersymmetric E$_{11}$ exceptional field theory,''
\doi{JHEP \textbf{10} (2019), 165}{doi:10.1007/JHEP10(2019)165}
\eprintN{1907.02080}.

\bibitem{West:2003fc}
P.~C.~West,
``E(11), SL(32) and central charges,''
\doi{Phys. Lett. B \textbf{575} (2003), 333-342}{doi:10.1016/j.physletb.2003.09.059}
\eprint{hep-th/0307098}.

\bibitem{Damour:2007dt}
T.~Damour, A.~Kleinschmidt and H.~Nicolai,
``Constraints and the E10 coset model,''
\doi{Class. Quant. Grav. \textbf{24} (2007), 6097-6120}{doi:10.1088/0264-9381/24/23/025}
\eprintN{0709.2691}.

\bibitem{Coimbra:2011ky}
A.~Coimbra, C.~Strickland-Constable and D.~Waldram,
``$E_{d(d)} \times \mathbb{R}^+$ generalised geometry, connections and M theory,''
\doi{JHEP \textbf{02} (2014), 054}{doi:10.1007/JHEP02(2014)054}
\eprintN{1112.3989}.

\bibitem{Berman:2012vc}
D.~S.~Berman, M.~Cederwall, A.~Kleinschmidt and D.~C.~Thompson,
``The gauge structure of generalised diffeomorphisms,''
\doi{JHEP \textbf{01} (2013), 064}{doi:10.1007/JHEP01(2013)064}
\eprintN{1208.5884}.

\bibitem{GGN} H. Godazgar, M. Godazgar and H. Nicolai,
``Einstein-Cartan Calculus for Exceptional Geometry",
\doi{JHEP \textbf{06} (2014), 021}{doi:10.1007/JHEP06(2014)021} 
\eprintN{1401.5984}.

\bibitem{KacMoodyWakimoto}
V.~G. Kac, R.~V. Moody, and M.~Wakimoto, ``On {$E_{10}$},'' in {\em
  Differential geometrical methods in theoretical physics ({C}omo, 1987)},
  vol.~250 of {\em NATO Adv. Sci. Inst. Ser. C Math. Phys. Sci.}, pp.~109--128.
\newblock Kluwer Acad. Publ., Dordrecht, 1988.

\bibitem{Moody:1968}
 R.~V.~Moody,
``A new class of Lie algebras,''
\doi{J. Algebra 10 (1968), 211--230}{10.1016/0021-8693(68)90096-3}.

\bibitem{KN1} A. Kleinschmidt and H. Nicolai,
``E(10) and SO(9,9) invariant supergravity",
\doi{JHEP \textbf{07} (2004), 041}{doi:10.1088/1126-6708/2004/07/041}
\eprint{hep-th/0407101}.

\bibitem{KN2} A. Kleinschmidt and H. Nicolai,
``IIB supergravity and E(10)",
\doi{Phys. Lett. B \textbf{606} (2005), 391-402}{doi:10.1016/j.physletb.2004.12.006}
\eprint{hep-th/0411225}.

\bibitem{KNP} A.~Kleinschmidt, H. Nicolai and J. Palmkvist,
``K(E9) from K(E10)",
\doi{JHEP \textbf{06} (2007), 051}{doi:10.1088/1126-6708/2007/06/051}
\eprint{hep-th/0611314}.

\bibitem{Obers:1999um}
N.~A.~Obers and B.~Pioline,
``Eisenstein series and string thresholds,''
\doi{Commun. Math. Phys. \textbf{209} (2000), 275-324}{doi:10.1007/s002200050022}
\eprint{hep-th/9903113}.

\bibitem{Damour:2000hv}
T.~Damour and M.~Henneaux,
``E(10), BE(10) and arithmetical chaos in superstring cosmology,''
\doi{Phys. Rev. Lett. \textbf{86} (2001), 4749-4752}{doi:10.1103/PhysRevLett.86.4749}
\eprint{hep-th/0012172}.

\bibitem{Damour:2002et}
T.~Damour, M.~Henneaux and H.~Nicolai,
``Cosmological billiards,''
\doi{Class. Quant. Grav. \textbf{20} (2003), R145-R200}{doi:10.1088/0264-9381/20/9/201}
\eprint{hep-th/0212256}.

\bibitem{Henneaux:2007ej}
M.~Henneaux, D.~Persson and P.~Spindel,
``Spacelike Singularities and Hidden Symmetries of Gravity,''
\doi{Living Rev. Rel. \textbf{11} (2008), 1}{doi:10.12942/lrr-2008-1}
\eprintN{0710.1818}.

\bibitem{Misner:1969hg}
C.~W.~Misner,
``Mixmaster universe,''
\doi{Phys. Rev. Lett. \textbf{22} (1969), 1071-1074}{doi:10.1103/PhysRevLett.22.1071}.

\bibitem{Graham:1990jd}
R.~Graham and P.~Szepfalusy,
``Quantum creation of a generic universe,''
\doi{Phys. Rev. D \textbf{42} (1990), 2483-2490}{doi:10.1103/PhysRevD.42.2483}.

\bibitem{Forte:2008jr}
L.~A.~Forte,
``Arithmetical Chaos and Quantum Cosmology,''
\doi{Class. Quant. Grav. \textbf{26} (2009), 045001}{doi:10.1088/0264-9381/26/4/045001}
\eprintGR{0812.4382}.

\bibitem{Kleinschmidt:2005gz}
A.~Kleinschmidt and H.~Nicolai,
``E(10) cosmology,''
\doi{JHEP \textbf{01} (2006), 137}{doi:10.1088/1126-6708/2006/01/137}
\eprint{hep-th/0511290}.

\bibitem{Bessel} see {\em e.g.}: \href{https://dlmf.nist.gov/10.9}{https://dlmf.nist.gov/10.9}.

\bibitem{Terras}
  A.~Terras,
  \doi{{\sl Harmonic analysis on symmetric spaces and applications I}}{https://doi.org/10.1007/978-1-4612-5128-6},
  Springer Verlag (New York, 1985).

\bibitem{Iwaniec} H.~Iwaniec, \doi{{\it Spectral methods of automorphic forms}}{http://dx.doi.org/10.1090/gsm/053}, 
  Am. Math. Soc. Graduate Studies in Mathematics Vol.~53 (2002).

\bibitem{Lax:1982} 
 P.~D.~Lax and R.~S.~Phillips, 
 ``The asymptotic distribution of lattice points in Euclidean and non-Euclidean spaces,'' 
 \doi{J. Functional Analysis \textbf{46} (1982) 280--350}{https://doi.org/10.1016/0022-1236(82)90050-7}.

\bibitem{Kazhdan:2001nx}
D.~Kazhdan, B.~Pioline and A.~Waldron,
``Minimal representations, spherical vectors, and exceptional theta series,''
\doi{Commun. Math. Phys. \textbf{226} (2002), 1-40}{doi:10.1007/s002200200601}
\eprint{hep-th/0107222}.

\bibitem{Gunaydin:2001bt}
M.~G\"unaydin, K.~Koepsell and H.~Nicolai,
``The minimal unitary representation of E(8(8)),''
\doi{Adv. Theor. Math. Phys. \textbf{5} (2002), 923-946}{doi:10.4310/ATMP.2001.v5.n5.a3}
\eprint{hep-th/0109005}.

\bibitem{Green:2011vz}
M.~B.~Green, S.~D.~Miller and P.~Vanhove,
``Small representations, string instantons, and Fourier modes of Eisenstein series,''
\doi{J. Number Theor. \textbf{146} (2015), 187-309}{doi:10.1016/j.jnt.2013.05.018}
\eprintN{1111.2983}.


\bibitem{Breitenlohner:1982jf}
P.~Breitenlohner and D.~Z.~Freedman,
``Stability in Gauged Extended Supergravity,''
\doi{Annals Phys. \textbf{144} (1982), 249}{doi:10.1016/0003-4916(82)90116-6}.

\bibitem{Moore} G.~Moore,
``Finite in all directions",
\eprint{hep-th/9305139}.

\bibitem{Weinberg} S.~Weinberg,
``Gravitation and Cosmology",
John Wiley and Sons (1972).

\bibitem{Halliwell}
J.~J.~Halliwell,
``Derivation of the Wheeler-De Witt Equation from a Path Integral for Minisuperspace Models,''
\doi{Phys. Rev. D \textbf{38} (1988), 2468}{doi:10.1103/PhysRevD.38.2468}.

\end{thebibliography}
\end{document}